\documentclass[10pt,journal]{IEEEtran}
\IEEEoverridecommandlockouts
\usepackage{amsmath,amssymb,amsfonts}
\usepackage{algorithmic}
\usepackage{graphicx}
\usepackage{textcomp}
\usepackage{xcolor,color}
\usepackage{float}
\usepackage{caption}
\usepackage{algorithm}
\usepackage[style=ieee, giveninits=true, backend=biber,doi=false]{biblatex}

\usepackage{etoolbox}
\AtBeginBibliography{\footnotesize}

\def\BibTeX{{\rm B\kern-.05em{\sc i\kern-.025em b}\kern-.08em
    T\kern-.1667em\lower.7ex\hbox{E}\kern-.125emX}}

\begin{document}
\title{Deep Learning-Empowered Movable-Antenna Position Optimization with Partial CSI}
\author{Lele~Lu,
        Weidong~Mei,~\IEEEmembership{Member,~IEEE},
        Xin~Wei,~\IEEEmembership{Student Member,~IEEE},
        Ruixi~Feng,
        Haocheng~Hua,~\IEEEmembership{Member,~IEEE},
        Zhi~Chen,~\IEEEmembership{Senior Member,~IEEE},
        Boyu~Ning,~\IEEEmembership{Member,~IEEE},
        and Emil Bj{\"o}rnson,~\IEEEmembership{Fellow,~IEEE}\vspace{-9pt}
\thanks{This paper was presented in part at the 2025 IEEE 36th International Symposium on Personal, Indoor and Mobile Radio Communications (PIMRC), Istanbul, Turkiye \cite{lu2025pimrc}.}
\thanks{L. Lu and R. Feng are with Glasgow College, University of Electronic Science and Technology of China, Chengdu, China (e-mail: ll.lu@std.uestc.edu.cn; 2022190905006@std.uestc.edu.cn).}%
\thanks{W. Mei, X. Wei, Z. Chen, and B. Ning are with the National Key Laboratory of Wireless Communications, University of Electronic Science and Technology of China, Chengdu, China (e-mail: wmei@uestc.edu.cn; xinwei@std.uestc.edu.cn; chenzhi@uestc.edu.cn; boydning@outlook.com).}%
\thanks{H. Hua is with the School of Science and Engineering, The Chinese University of Hong Kong (Shenzhen), Guangdong 518172, China (e-mail: huahaocheng@cuhk.edu.cn).}%
\thanks{E. Bj{\"o}rnson is with the Department of Communication Systems, KTH Royal Institute of Technology, Stockholm, Sweden (email: emilbjo@kth.se).}
}
\maketitle

\begin{abstract}
Movable antennas (MAs) have emerged as a promising technology to achieve high data rates in wireless communications by dynamically adjusting their positions to mitigate deep fading within a given region. However, to determine the optimal MA positions, full channel state information (CSI) is required for each antenna position within the transmit/receive movement region, which leads to extremely high channel estimation overhead. To tackle this challenge, this paper proposes a deep neural network (DNN)-based learning framework to predict the optimal positions of multiple transmit MAs in a multi-user multiple-input single-output (MISO) system without explicit channel estimation. To unveil useful insights, we first consider a simpler single-user MISO case, showing that there exists a clear mapping between the optimal MA positions and the channel power gains from a subset of locations within the transmit region to the user. However, this mapping is highly nonlinear and cannot be explicitly characterized for practical channel models. To tackle this challenge, we train a deep neural network to learn it in a supervised manner and then use the pre-trained DNN to determine the optimized MA positions in real-time data transmission, based on partial power measurements within the transmit region only. However, this framework cannot be applied to the multi-user case due to the more complex rate expression and the unavailability of globally optimal antenna position solutions as labels. To tackle this difficulty, we develop an unsupervised training framework to directly maximize the multi-user sum-rate. In particular, an attention-based architecture is employed to extract latent features from partial channel measurements and manage inter-user interference. Simulation results demonstrate that the proposed framework achieves near-optimal performance in single-user systems and even outperforms conventional CSI-based alternating optimization algorithms in the multi-user case. 
\end{abstract}

\section{Introduction}
Movable antenna (MA) technology has recently emerged as a compelling means to enhance the performance of wireless communications by allowing antennas to be flexibly repositioned within a prescribed movement region at the transmitter and/or receiver \cite{Zhu2025,Zhu2023,Ning2025}. In contrast to conventional fixed-position antennas (FPAs), MAs can proactively improve channel conditions by avoiding fading and making user channels nearly orthogonal, thereby offering improved reliability and spectral efficiency. Beyond throughput gains, the geometric reconfigurability of MAs has also shown promise for wireless sensing \cite{ma2024movable,wang2025sensing} and array signal processing \cite{ma2024multi,zhu2023movable,wang2026movable,wang2026filter}, underscoring their potential for future wireless networks.

The promising and multifarious benefits of movable antennas (also known as fluid antennas from the viewpoint of flexible antenna positions \cite{New2024}) have spurred extensive research interest.
\begin{figure*}[!t]
  \centering
  \includegraphics[width=0.9\textwidth]{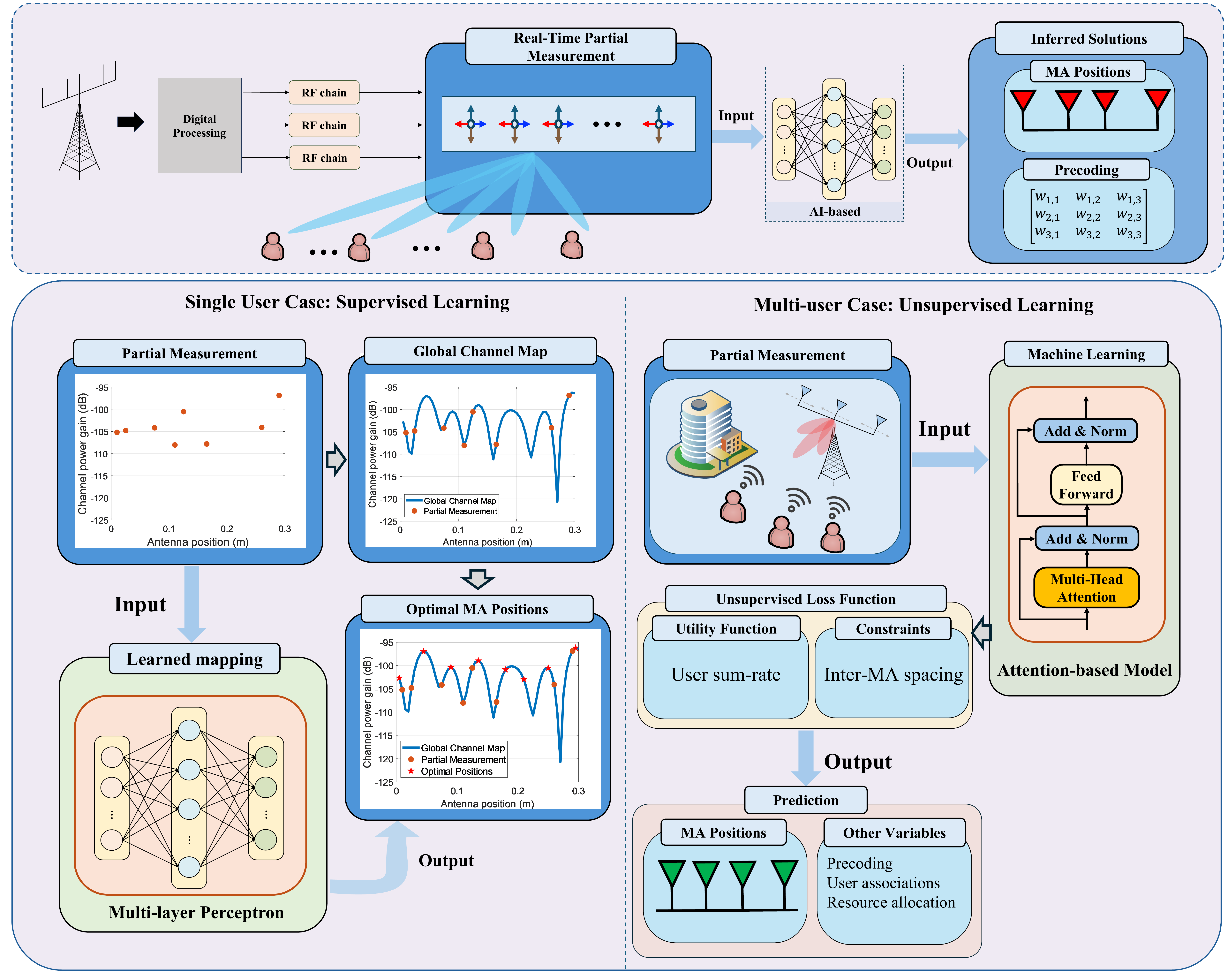}
  \caption{System model and illustration of the proposed learning-based frameworks.}
  \label{model}
  \vspace{-6pt}
\end{figure*}
A central line of research has focused on MA position optimization techniques in view of the highly nonlinear multi-path channel expressions in terms of antenna positions. To tackle this challenge, some existing works consider optimizing MA positions based on long-term or statistical channel characteristics, where antenna locations are designed offline to achieve favorable average performance under given deployment scenarios. For example, MA position optimization based on statistical channel state information (CSI) was investigated in \cite{Yan2025StatCSI}, where antenna positions require less frequent updates compared to those based on instantaneous CSI. In addition, the authors in \cite{irshad2025pre} proposed a pre-optimized irregular array architecture, where antenna geometry is optimized for a given deployment scenario through offline performance evaluation, which dispenses with instantaneous CSI as well. Although these approaches can significantly reduce the overhead of channel training and online optimization, their real-time performance may degrade, particularly in time-varying environments with dynamic channel fluctuations.

For real-time antenna position optimization, most existing works developed a variety of gradient-based algorithms for antenna position optimization based on a theoretical field-response channel model \cite{Zhu2025}. For example, the authors in \cite{ma2024mimo} proposed an alternating optimization (AO) algorithm to maximize the capacity of an MA-enhanced multiple-input multiple-output (MIMO) system, where the successive convex approximation technology is employed to optimize the transmit and receive antenna positions. Moreover, the authors in \cite{hu2024secure,Wang2024a,Zhu2023b,li2025relay} applied gradient-based algorithms for antenna position optimization in the scenarios of physical-layer security, interference networks, multi-user uplink transmission, and relay systems, respectively.

However, due to the highly nonlinear channel expression, gradient-based algorithms are sensitive to initialization and may converge to poor local optima. To enhance global search capability, a number of works have resorted to meta-heuristic and nature-inspired optimization methods that do not rely on gradient information. For example, the authors in \cite{Hoang2024FireflyMA} proposed a firefly algorithm to optimize MA positions, inspired by brightness-driven population interactions. Particle swarm optimization (PSO) and its variants have also been applied to MA systems, e.g., for multi-user communication systems \cite{xiao2024multiuser}. While these heuristic algorithms are more capable of escaping poor local minima, they generally incur high computational complexity and lack theoretical guarantees on even local optimality, which limits their scalability in practical MA deployments.

To mitigate these issues, an alternative line of research discretizes the movement region into a set of candidate sampling points, thereby converting the continuous MA position optimization into a discrete selection problem. This reformulation facilitates more efficient solution methods based on discrete optimization algorithms. For example, in \cite{Mei2024}, a graph-based method was developed to solve the discrete point selection {\it optimally} in polynomial time for MA-enhanced multiple-input single-output (MISO) systems. For multi-user MA-MISO systems, the authors in \cite{Wu2023} obtained the globally optimal MA positions via a generalized Bender's decomposition (GBD) algorithm. The discrete MA position optimization has also been applied for physical-layer security \cite{mei2024posistion,shen2025movable}, cognitive radio \cite{wei2024joint}, intelligent reflecting surface (IRS)-aided communications \cite{wei2025irs}, robust antenna position optimization \cite{ma2025robust}, MA trajectory designs \cite{li2026traj}, orthogonal frequency division multiplexing (OFDM) \cite{Feng2025MISOofdm}, among others. More recently, a general and efficient framework was established in \cite{Liu2026DiscreteFramework} to unify discrete MA optimization by combining the sequential update and Gibbs sampling techniques.

However, to practically implement the above antenna position optimization algorithms, a crucial question lies in how to acquire the CSI across the entire antenna movement region, referred to as the channel map. Compared with the channel estimation in conventional FPA systems, the training and feedback overhead becomes significantly higher. To deal with this issue, compressed sensing-based methods were proposed in \cite{ma2023compressed,xiao2024channel} to recover the channel map by leveraging the structural properties of the field-response channel model. To further improve the accuracy of channel map reconstruction, tensor decomposition techniques were introduced in \cite{zhang2024channel} by leveraging structured pilot measurements. While effective in channel map estimation, these approaches remain dependent on a specific channel model, which can introduce model-mismatch issues in practical deployments.

To improve the robustness of the above model-based channel map methods, data-driven approaches have recently attracted increasing attention. For example, Bayesian reconstruction methods \cite{zhang2025successive} and minimum mean square error (MMSE)-based estimation schemes \cite{Skouroumounis2022} exploit statistical prior information or spatial correlation to enhance channel inference robustness. In addition, correlation-based deep learning techniques were developed in \cite{Ji2024,huang2025} for channel map estimation in MA systems via channel extrapolation \cite{Ji2024,huang2025}. However, most existing learning-based methods still aim to reconstruct the channel maps and then optimize the MA positions, which may overlook an implicit direct mapping between the optimal MA positions and partial CSI within the antenna movement region. Intuitively, the optimal positions are determined by the full channel map. Yet, within a fixed environment, channel responses at different positions are not independent but are governed by the same scatterers. Consequently, CSI measured at a subset of positions can carry latent information about the full channel map (and thus about the optimal MA positions).

Motivated by these observations, this paper proposes a deep learning–empowered framework for MA position optimization that infers high-quality MA placements using channel measurements from only a subset of sampling points. By bypassing explicit channel reconstruction, the proposed method enables efficient MA position optimization with substantially reduced channel training overhead and significantly lower optimization complexity. Specifically, this paper focuses on the downlink of a multi-user MISO communication system, where a multi-MA base station (BS) communicates with multiple single-antenna users at the same time, as shown in Fig.~\ref{model}. The main contributions of this paper are summarized as follows.
\begin{itemize}
    \item To unveil insights, we first consider a simpler single-user MISO case. We show that a well-defined mapping exists between the optimal MA positions and the channel power gains measured at only a subset of locations within the transmit region, although this mapping is difficult to characterize analytically. To address this issue, we propose a up-down multi-layer perception (MLP) architecture to approximate the mapping via offline supervised learning, using the globally optimal MA positions (computed using the graph-based algorithm in \cite{Mei2024}) as ground-truth labels, as illustrated in Fig.~\ref{model}. After training, the model is deployed for real-time inference, which directly outputs the MA positions from partial power measurements within the transmit region.
    \item However, this supervised learning framework cannot be directly extended to the multi-user case due to the more involved sum-rate expression and the absence of globally optimal antenna-position labels. To tackle this difficulty, we develop an unsupervised learning framework that directly maximizes the multi-user sum-rate, as illustrated in Fig.~\ref{model}. Specifically, we propose an attention-based architecture that extracts latent features from partial CSI measurements, manages inter-user interference, and aggregates per-user representations through three dedicated modules, respectively. By incorporating the sum-rate objective and the associated constraints into the loss function, the model learns to produce high-quality MA positioning (and transmit beamforming) solutions without explicit supervision. Simulation results validate the efficacy of the proposed learning-based methods in both single- and multi-user setups.
\end{itemize}

The remainder of this paper is organized as follows. Section~II presents the system model and problem formulation. Sections~III and IV present the proposed supervised and unsupervised learning frameworks for the single- and multi-user cases, respectively. Section~V presents simulation results. Section~VI concludes the paper.

\textbf{Notations:} Bold lowercase and uppercase letters (e.g., $\mathbf{h}$ and $\mathbf{H}$) denote vectors and matrices, respectively. Superscripts $(\cdot)^{\top}$ and $(\cdot)^{H}$ represent transpose and Hermitian transpose, respectively. $\|\cdot\|_2$ and $\|\cdot\|_F$ are the Euclidean and Frobenius norms, respectively. The complex and real domains are denoted by $\mathbb{C}$ and $\mathbb{R}$, respectively. $\mathbb{E}[\cdot]$ denotes expectation. 

\begingroup
\allowdisplaybreaks
\section{System Model}
As shown in Fig.\,\ref{model}, we consider a narrowband MISO communication system where an $N$-MA BS communicates with $K$ single-FPA users. Let $\mathcal{K} = \{1,2,\cdots, K\}$ denote the set of the $K$ users. The $N$ transmit MAs can be flexibly moved within a one-dimensional linear array with a length of \(A\) in meters (m). Let \(\mathcal{N} = \{1, 2, \cdots, N\}\) denote the set of all MAs. We assume that the channels between the BS and all users are quasi-static (e.g., for smart homes and factories), such that the MAs can move to their optimized positions within the channel coherence time. Considering the practical finite-resolution constraints of antenna position adjustment (e.g., those imposed by stepper motors or discrete architectures such as pixel antennas), we uniformly sample the MA array into \(M\) (\(M \gg N\)) discrete positions, with the distance between any two adjacent sampling points given by \(\delta_s = \frac{A}{M}\). Therefore, the position of the \(m\)-th sampling point can be represented as \(s_m = \frac{mA}{M}, m \in \mathcal{M} = \{1, 2, \dots, M\}\), and the position of each MA can be selected from one of the sampling points in \(\mathcal{M}\).

Let \(a_n\) denote the index of the sampling point for the $n$-th MA. Hence, the position of this antenna is \(s_{a_n} = \frac{a_n A}{M}, \quad n \in\mathcal{N}\). To avoid mutual coupling between MAs, we introduce a minimum spacing between any two MAs, denoted as \(d_{\text{min}}\). Therefore, we have
\begin{equation}\label{dist}
    \lvert a_i - a_j \rvert \geq a_{\text{min}}, \quad \forall i, j \in \mathcal{N}, \, i \neq j.
\end{equation}
where \(a_{\text{min}} = d_{\text{min}}/{\delta_s} \gg 1\). Thus, optimizing the MA positions is equivalent to selecting sampling points from the set \(\mathcal{M}\) subject to the minimum spacing constraint in  \eqref{dist}. Here, we define an antenna index vector (AIV) $\mathbf{a}=\begin{bmatrix} a_{1} & a_{2} & \cdots & a_{N}\end{bmatrix}^\top$ to represent the selected antenna positions.

Let \(h_{k}^{a_n}\in\mathbb{C}\) be the complex baseband channel from the $n$-th BS antenna to user $k$, \(n\in\mathcal{N}\),  \(k\in\mathcal{K}\). For any given AIV, the downlink channel vector for user \(k\) is
\begin{equation}
  \mathbf{h}_k
  = \begin{bmatrix} h_{k}^{a_1} & h_{k}^{a_2} & \cdots & h_{k}^{a_N} \end{bmatrix}^{\top}
  \in\mathbb{C}^{N\times 1}.
\end{equation}
Then, the aggregated BS–user channel matrix is written as
\begin{equation}\label{channelmap}
    \mathbf{H}(\mathbf{a}) =
    \begin{bmatrix}
        h_1^{a_1} & h_1^{a_2} & \cdots & h_1^{a_N} \\
        h_2^{a_1} & h_2^{a_2} & \cdots & h_2^{a_N} \\
        \vdots & \vdots & \ddots & \vdots \\
        h_K^{a_1} & h_K^{a_2} & \cdots & h_K^{a_N}
    \end{bmatrix}
    \in \mathbb{C}^{K \times N}.
\end{equation}
\; Let \(\mathbf{w}_k\in\mathbb{C}^{N\times 1}\) be the precoding vector for user \(k\) with $\sum_{k \in \mathcal{K}}\|\mathbf{w}_k\|^2_2 \leq P_t$, where $P_t$ is the BS's transmit power. The received signal at user \(k\) is given by
\begin{equation}\label{sig}
  y_k = \mathbf{h}_k^{H}(\mathbf{a})\mathbf{w}_k x_k
      + \sum_{i\neq k}\mathbf{h}_k^{H}(\mathbf{a})\mathbf{w}_i x_i
      + n_k,
\end{equation}
where $x_k$ is the transmitted data symbol for user $k$ with \(\mathbb{E}[|x_k|^2]=1\),  and \(n_k\sim\mathcal{CN}(0,\sigma_n^2)\) represents the received noise at user $k$ with $\sigma^2_n$ denoting its average power.  Based on \eqref{sig}, by treating interference as noise, the achievable rate at user $k$ is given by
\begin{equation}
  R_k(\mathbf{W},\mathbf{a})
  = \log_2\!\left(
      1+\frac{\big|\mathbf{h}_k^{H}(\mathbf{a})\mathbf{w}_k\big|^2}
                {\sum\limits_{i\neq k}\big|\mathbf{h}_k^{H}(\mathbf{a})\mathbf{w}_i\big|^2+\sigma^2_n}
    \right),
\end{equation}
where \(\mathbf{W}=[\mathbf{w}_1,\ldots,\mathbf{w}_K]\in\mathbb{C}^{N\times K}\) denotes the transmit precoding matrix. Our objective is to maximize the sum-rate of the $K$ users\footnote{It is worth noting that the proposed algorithms are also applicable to other performance metrics, such as the common achievable rate among all users.} by jointly optimizing the BS's transmit precoding matrix $\mathbf{W}$ and the positions of MAs. Hence, the optimization problem is formulated as
\begin{subequations}\label{P1}
\begin{align}
  \text{(P1)}\;\max_{\mathbf{W},\,\{\mathbf{a}\}}~~
    & R_{\mathrm{sum}}(\mathbf{W},\mathbf{a}) = \sum_{k=1}^{K} R_k(\mathbf{W},\mathbf{a})\notag \\[1mm]
  \text{s.t.}\quad
    & \sum_{k=1}^{K}\|\mathbf{w}_k\|_2^2 \le P_t,
    \label{P1a}\\
    & a_n \in \mathcal{M},\quad \forall n\in\mathcal{N}, \label{P1b}\\
    & |a_i-a_j| \ge a_{\min},\quad \forall i \neq j,~ i,j\in\mathcal{N}. \label{P1c}
\end{align}
\end{subequations}
However, (P1) is a non-convex optimization problem that is challenging to solve due to the inter-MA spacing constraints (i.e., \eqref{P1c}). Moreover, solving (P1) requires channel maps associated with all $K$ users, i.e., $h_k^m, m \in {\cal M}, k \in {\cal K}$ involving $KM$ parameters, which may incur high channel estimation overhead in practice. To reveal essential insights and facilitate the exposition of our proposed partial CSI-based design, we first consider a simplified case with a single user and propose a learning-based solution accordingly. The extension to the general multi-user case will be presented in Section IV.\vspace{-6pt}

\section{Single-User Case}
In this section, we consider a special single-user scenario for (P1). In this scenario, to maximize the user’s achievable rate, the maximum ratio transmission (MRT) should be adopted to maximize the received signal power at the user in the absence of inter-user interference, i.e.,
\begin{equation} \label{wt}
    \mathbf{w} = \frac{\sqrt{P_t}}{\| \mathbf{h}(\mathbf{a}) \|} \mathbf{h}(\mathbf{a}).
\end{equation}
The resulting maximum received signal power is given by
\begin{equation} \label{pr}
    P_r(\mathbf{a}) = |\mathbf{w}^H \mathbf{h}(\mathbf{a})|^2 = P_t \sum_{n=1}^N |h_{a_n}|^2.
\end{equation}
where $\mathbf{w} \in \mathbb{C}^N \times 1$ denotes the transmit beamforming with $\|\mathbf{w}\|_2^2 = P_t$,  and the subscripts “$k$” in $\mathbf{h}_k$ and $\mathbf{w}_k$ are omitted here for brevity. 
\subsection{Existing CSI-Based Solution}
Let $\mathcal{H}=\{h_m\}_{m \in \cal M}$ denote the channels from all sampling points in $\cal M$ to the user, i.e., channel map. In the case that this map is available (e.g., via the channel estimation techniques proposed for MAs in \cite{ma2023compressed,xiao2024channel,zhang2024channel,zhang2025successive}), the optimal MA positions can be obtained by solving the following optimization problem, i.e.,
\begin{subequations}\label{P2}
    \begin{align}
    \text{(P2)}\; \max_{\mathbf{a}}\; & \sum_{n=1}^{N} |h_{a_n}|^2, \notag\\[1mm]
    \text{s.t.} \;& a_n \in \mathcal{M}, \quad \forall n \in \mathcal{N}, \label{P2a}\\
    & \lvert a_i - a_j\rvert \ge a_{\min}, \;\forall\, i,j \in \mathcal{N},\; i\neq j,\label{P2b}
    \end{align}
\end{subequations}
where we have omitted the constant $P_t$ in the objective function, as it does not affect the solution. 

Although (P1) is a combinatorial optimization problem that is generally difficult to optimally solve, in our previous work \cite{Mei2024}, we have proposed a graph-based algorithm to solve (P1) optimally in polynomial time. Specifically, a directed weighted graph $G=(V_0,E_0)$ is constructed, where $V_0$ and $E_0$ denote its vertex and edge sets, respectively. Moreover, we model each sampling point as a vertex in $G$, i.e., $V_0={\cal M}$. To construct the set of edges, an edge is added between two vertices $i$ and $j$ if their corresponding sampling points satisfy \eqref{dist}, i.e., $\lvert i-j \rvert \ge a_{\min}$. Finally, by properly assigning each edge with a weight based on ${\cal H}$, it can be shown that (P1) is equivalent to finding an $(N+1)$-hop shortest path in $G$. The details are omitted for brevity. This fixed-hop shortest path problem can be optimally solved by applying dynamic programming.

However, the graph-based solution relies on the channel map ${\cal H}$, which can be practically difficult to acquire, especially in the complex rich-scattering environment with a large transmit array. Fortunately, the proposed solution in \cite{Mei2024} reveals the existence of a clear mapping from ${\cal H}$ to the optimal MA positions. Moreover, CSI at different positions within the linear array often exhibits a certain degree of correlation due to the common environmental scatterers. This implies that the channel map can be inferred from CSI at partial positions within the transmit array \cite{Skouroumounis2022,zhang2024channel,huang2025, Ji2024}. Consequently, a mapping may exist between the CSI at these partial positions and the optimal MA positions. Motivated by this, we propose a more efficient learning-based approach that circumvents explicit CSI estimation in the following.\vspace{-6pt}

\subsection{Proposed Optimization Framework with Partial CSI}
In this section, we present our proposed learning-based framework for optimizing MA positions. Note that in the considered single-user case with MRT, it suffices to determine the optimal MA positions based on the channel power gain at each position, i.e., $\lvert{h}_{m}\rvert^2,  m \in {\cal M}$. As such, we only measure partial BS-user channel power gains at $S$ training sampling points with \(S \ll M\). To this end, we assume a time division duplex (TDD) system in this paper, such that the downlink channel power gains can be measured by estimating their uplink counterparts based on the assumption of channel reciprocity.\footnote{In the frequency division duplex system, the downlink channel power gains can also be measured based on the feedback from the user over control links.} The user then sends a unit-power beacon signal to the BS. Denote by $\boldsymbol{y} = [y(1), y(2), \dots, y(S)]^\top \in \mathbb{R}^{S \times 1}$ the beacon signal power received over the $S$ training sampling points, which is used to approximate the channel power gains over these points in the presence of noise. Let
\begin{equation}
f(\cdot): \mathbb{R}^{S \times 1} \rightarrow {\cal M}^N
\end{equation}
denote the mapping from $\boldsymbol{y}$ to the optimal MA positions.

Given the vector $\boldsymbol{y}$ and the mapping $f(\cdot)$, our goal is to optimize the MA positions by solving the following optimization problem, i.e.,
\begin{align}
\text{(P2)} \;\max_{\{a_n\} = f(\boldsymbol{y})} \sum_{n=1}^{N} |h_{a_n}|^2, \;\;
\text{s.t.}\; {\text{\eqref{P1a},\eqref{P1b}}}.
\end{align}
Compared to (P1), solving (P2) dispenses with the estimation of the channel map ${\cal H}$ and only relies on $\boldsymbol{y}$. However, it is still challenging to solve (P2) due to the implicit expression of the mapping $f(\cdot)$, which is highly complex to characterize. To deal with this challenge, we propose a deep neural network (DNN) that learns to predict optimal antenna positions by approximating $f(\cdot)$.
\begin{figure}[!t]
  \centering
  \includegraphics[clip, page =3, width=1\linewidth]{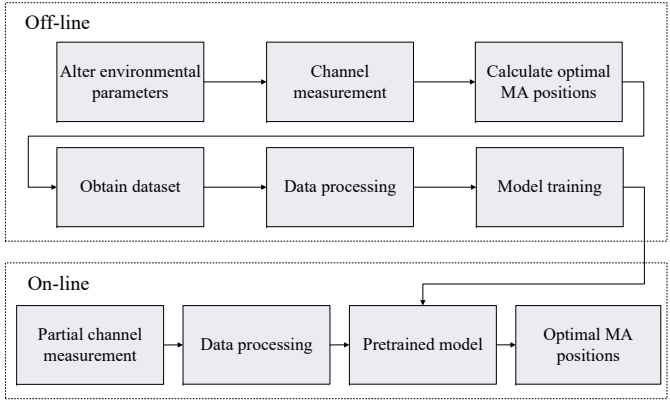}
  \caption{Main steps of the proposed learning-based method.}
  \label{fig:process}
\end{figure}

\begin{figure*}[t]
  \centering
  \includegraphics[clip, page = 1, width=0.85\linewidth]{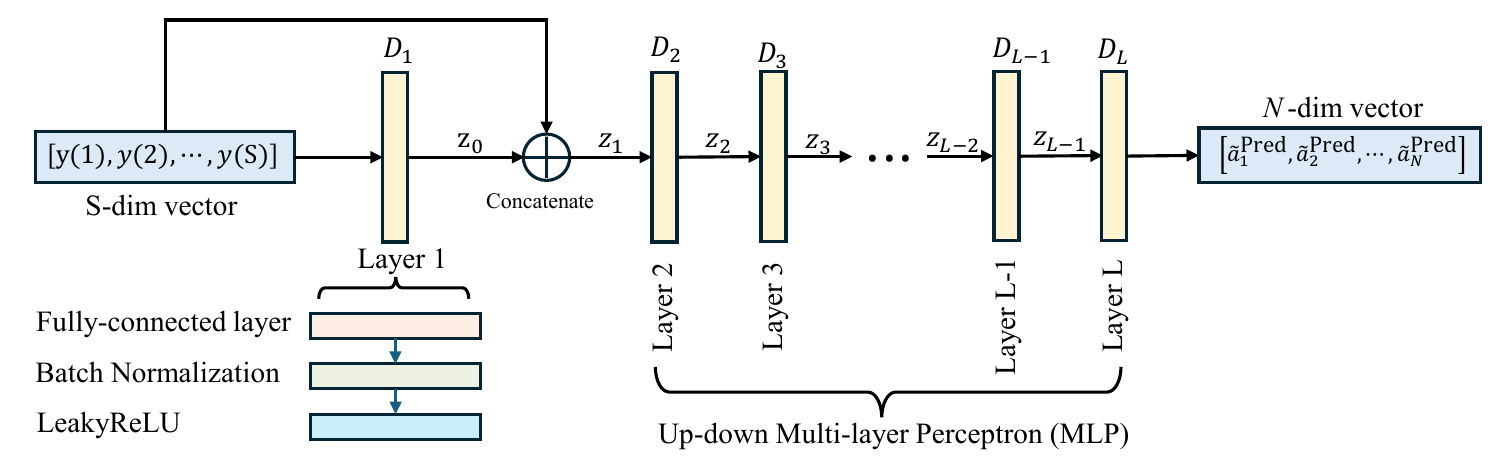}
  \caption{Proposed DNN architecture in the single-user case.}
  \label{fig:art}
  \vspace{-6pt}
\end{figure*}
To train the DNN, we design a two-stage optimization framework with hybrid offline and online processing, as depicted in Fig.~\ref{fig:process}. In the offline stage, we collect sufficient channel maps and their corresponding optimal MA positions via traditional approaches as presented in Section III-A, and then train the DNN in a supervised manner. In the real-time transmission,  $\boldsymbol{y}$ is input to the pre-trained DNN to obtain the optimized antenna positions directly (and the corresponding MRT beamformer), without the need to estimate $\mathcal{H}$. In the next section, we discuss the training process in detail.\vspace{-6pt}

\subsection{DNN Training}
The procedures of DNN training consist of the following key steps, namely, data collection, data processing, model design, and model training, as presented below.

\subsubsection{Offline Data Collection and Processing}
To generate a rich dataset for training in the offline phase, we explore scenarios with varying environmental parameters, such as the user's position, the number of multi-path components, the path-loss exponent, the reference path loss at a given distance, and the scatterer locations, etc. Let $T$ denote the total number of channel realizations by changing the above environmental parameters and $\mathcal{\tilde{H}}_t=\{\lvert\tilde{h}_{t,m}\rvert^2, m\in\mathcal{M}\}$, $t=1,2,\cdots,T$ denote the set of channel power gains at all sampling points in the $t$-th channel realization, where $\tilde{h}_{t,m}$ represents the channel from the $m$-th sampling point to the user in this realization. This channel power gain map can be practically estimated by applying the existing model-based or model-free algorithms for MAs \cite{Skouroumounis2022, zhang2024channel, Ji2024} or performing exhaustive channel estimation for a small length of linear array. Alternatively, we can generate the channel power gain map based on simulation if an accurate channel model (e.g., ray-tracing and electromagnetic simulation) is available for the considered environment. Then, we derive the indices of the optimal sampling points for the $t$-th channel realization, denoted as $\{a_{t,n}\}$, by applying the optimal graph-based algorithm in our previous work \cite{Mei2024}.

To ensure effective learning, all estimated channel power gains should be normalized. Specifically, in the $t$-th channel realization, the normalization is performed as
\begin{equation}
g_{t,m}=\frac{\lvert\tilde{h}_{t,m}\rvert^2-\mu_t}{\sigma_t},\,\,m\in\mathcal{M},
\end{equation}
where $\mu_t\!=\!\frac{1}{M} \sum_{m\in\mathcal{M}}{\lvert\tilde{h}_{t,m}\rvert^2}$ and $\sigma_t^2=\frac{1}{M} \sum_{m\in\mathcal{M}}{\lvert\tilde{h}_{t,m}\rvert^4}-\mu_t^2$ denote the mean value and variance for all channel power gains in $\mathcal{\tilde{H}}_t$, respectively. 
This normalization ensures that each set of CSI data has zero mean and unit variance, which helps handle the data with a wide range of values and mitigate potential training instabilities arising from large numerical variations.

\subsubsection{Model Design and Training}
The proposed neural network is built upon an up-down MLP architecture, as illustrated in Fig.~\ref{fig:art}. Denote by $L$ the total number of layers and $D_l$ the dimensionality of features at layer $l$.

The network takes as input the estimated channel power gains at $S$ training sampling points, denoted by $\boldsymbol{y} \in \mathbb{R}^{S \times 1}$. Initially, the first fully connected layer transforms $\boldsymbol{y}$ to extract informative high-level features, i.e.,
\begin{equation}
    \mathbf{z}_0 = \sigma\Big(\boldsymbol{W}_1 \boldsymbol{y} + \mathbf{b}_1\Big), \quad \boldsymbol{W}_1 \in \mathbb{R}^{(D_2-S) \times S},
\end{equation}
where $\sigma(\cdot)$ is the LeakyReLU activation function, and $\boldsymbol{W}_1$ and $\mathbf{b}_1$ are the weight matrices and bias vectors, respectively.

To effectively leverage both the raw measurements and the extracted features, the network concatenates the original input $\boldsymbol{y}$ with the high-level feature vector $\mathbf{z}_0$. This process  preserves the low-level signal characteristics present in $\boldsymbol{y}$ while simultaneously incorporating the nonlinear transformations captured by $\mathbf{z}_0$, resulting in a comprehensive feature representation \(\mathbf{z}_1\),
\begin{equation}
    \mathbf{z}_1 = \big[\boldsymbol{y};\, \mathbf{z}_0\big] \in \mathbb{R}^{D_2\times 1}.
\end{equation}

Subsequently, $\mathbf{z}_1$ is fed into layer 2, and the DNN enters an expansion phase where each successive layer doubles the feature dimensionality, i.e.,
\begin{equation}
    \dim(\mathbf{z}_{l+1}) = 2 \cdot \dim(\mathbf{z}_l), \;1 \le l \le L/2-1.
\end{equation}
where $\dim(\mathbf{z}_l)$ denotes the dimension of features at layer $l$. The maximum dimension is reached at layer $L/2$.

Upon reaching the peak dimension, the architecture transitions into a compression phase, in which the feature dimensionality is exponentially reduced through consecutive fully connected layers, i.e.,
\begin{equation}
    \dim(\mathbf{z}_{l+1}) = \frac{1}{2} \cdot \dim(\mathbf{z}_l), L/2 \le l \le L-1.
\end{equation}
Ultimately, the output of layer L is the optimized antenna position indices \(\{a_n\}\).
At the $l$-th layer, the transformation is carried out as
\begin{equation}
    \mathbf{z}_{l+1} = \sigma\left(\text{BN}\left(\boldsymbol{W}_{l+1}\mathbf{z}_l + \mathbf{b}_{l+1}\right)\right), 2 \le l \le L-1,
    \label{eq:layer_update}
\end{equation}
where batch normalization (BN) is applied to the outputs of each layer and given by
\begin{equation}
    \text{BN}(\mathbf{z}) = \boldsymbol{\gamma} \odot \frac{\mathbf{z} - \boldsymbol{\mu}_{\mathcal{B}}}{\sqrt{\boldsymbol{\sigma}_{\mathcal{B}}^2 + \varepsilon}} + \boldsymbol{\beta}.
\end{equation}
Here, $\boldsymbol{\mu}_{\mathcal{B}} \in \mathbb{R}^{D_l \times 1}$ and $\boldsymbol{\sigma}_{\mathcal{B}}^2 \in \mathbb{R}^{D_l \times 1}$ denote the channel-wise mean and variance computed over mini-batches, and $\varepsilon$ is a small positive constant introduced for numerical stability (e.g., $1\times 10^{-5}$). The symbols $\boldsymbol{\gamma}, \boldsymbol{\beta} \in \mathbb{R}^{D_l \times 1}$ represent learnable scaling and shifting parameters, respectively, while $\odot$ denotes the element-wise product between vectors. In this paper, we set the number of layers as $L=12$.

\subsubsection{Loss Function}
To guide the training process, we employ the mean squared error (MSE) as the loss function. Specifically, let $\tilde{a}_{t,n}$ denote the predicted index of the sampling point for the $n$-th MA in the $t$-th channel realization. Then, the loss function is given by
\begin{equation}
\text{MSE} = \frac{1}{T}\sum_{t=1}^{T}\sum_{n=1}^{N}({a}_{t,n}-\tilde{a}_{t,n})^2.
\end{equation}
By minimizing this MSE, the model is encouraged to produce predictions that closely match the ground truth, thus improving antenna positioning accuracy and overall performance in practical scenarios.

\subsubsection{Minimum Inter-MA Spacing}
It should be mentioned that in the proposed DNN, the constraint on the minimum spacing between MAs \eqref{P1b} is not explicitly enforced. However, through extensive training on ground-truth data, i.e., ${a}_{t,n}$'s, the pretrained model inherently learns to adhere to this constraint in its high-dimensional representation. Our numerical results confirm that violations are rare, and when they do occur, only minor adjustments are required to ensure compliance, as discussed next.

Specifically, let \(\{{\tilde a}^{\text{Pred}}_1, {\tilde a}^{\text{Pred}}_2, \dots, {\tilde a}^{\text{Pred}}_N\}\) denote the indices of MAs output by the DNN, with ${\tilde a}^{\text{Pred}}_1 < {\tilde a}^{\text{Pred}}_2 < \cdots < {\tilde a}^{\text{Pred}}_N$. To adjust these indices to meet \eqref{P1b}, we aim to generate a new sequence of MA indices, denoted as \(\{b_1, b_2, \dots, b_N\}\), that is feasible to (P2) and has the minimum difference from \(\{{\tilde a}^{\text{Pred}}_1, {\tilde a}^{\text{Pred}}_2, \dots, {\tilde a}^{\text{Pred}}_N\}\). This can be achieved by solving the following optimization problem, i.e.,
\begin{subequations}
    \begin{align}
    \text{(P3)}\quad \min_{\{b_n\}} \quad & \sum_{n=1}^{N} |b_n - {\tilde a}^{\text{Pred}}_n|, \notag\\[1mm]
    \text{s.t.} \quad & b_n \in \mathcal{M}, \quad \forall\, n \in \mathcal{N}, \label{P3a}\\[1mm]
    & \lvert b_i - b_j\rvert \ge a_{\min}, \quad \forall\, i,j \in \mathcal{N},\; i\neq j, \label{P3b}
    \end{align}
\end{subequations}
where we use $\lvert b_n - {\tilde a}^{\text{Pred}}_n \rvert$ to characterize the difference between $b_n$ and ${\tilde a}^{\text{Pred}}_n, n \in \cal N$.

Although (P3) is a discrete optimization problem, its optimal solution can be obtained by applying a dynamic programming algorithm, which recursively computes the optimal value of (P3). Specifically, in the case of $N=n$, we define \(A_{n,m}\) and \(I_{n-1,m}\) as the optimal value of (P3) and the optimal index of the \((n-1)\)-th MA if we set $b_n=m$, respectively. For $N=1$, it is obvious to see
\begin{equation}
    A_{1,m} = |m - a^{\text{Pred}}_1|, \quad m = 1, 2, \dots, M.
\end{equation}
Currently, \( I_{0,m} \) can be left empty or assigned an initial marker.

Then, we can recursively derive $A_{n,m}$ from $A_{n-1,m}$ based on the following relation,
\begin{equation}
    A_{n,m} = \min_{j:\; j \leq m -  a_{\min}} A_{n-1,j} + |m - a^{\text{Pred}}_n|,
\end{equation}
where we constrain the index of the $(n-1)$-th MA no larger than $m -  a_{\min}$ to meet the minimum distance constraint with the $n$-th MA. The optimal index of the $(n-1)$-th MA under $b_n=m$ is thus given by
\begin{equation}
   I_{n-1,m} = \arg\min_{j:\; j \leq m - a_{\min}} A_{n-1,j}.
\end{equation}
Finally, the optimal index of the $N$-th MA is determined as
\begin{equation}
    b_N^* = \arg\min_{m \in \cal M} A_{N,m},
\end{equation}
and the indices of all previous $N-1$ MAs can be recursively obtained as
\begin{equation}
    b^*_{n-1} = I_{n-1,b^*_n}, \; 2 \leq n \leq N.
\end{equation}
Based on the above, the optimized AIV by the proposed learning-based algorithm can be obtained.

\section{Multi-User Case}
In this section, we consider the general multi-user case and solve (P1) accordingly. Similar to Section III, we first review the existing CSI-based solution to (P1).\vspace{-6pt}

\subsection{Existing CSI-Based Solution}\label{multiUserAO}
In the presence of channel maps for all users, (P1) can be efficiently solved by an AO-based algorithm. In particular, (P1) is decomposed into two subproblems: (i) transmit precoding matrix optimization at the BS for a given AIV, and (ii) AIV optimization for a given transmit precoding matrix.

\subsubsection{Optimizing $\mathbf{W}$ with Given ${\mathbf a}$}
First, with any given AIV ${\mathbf a}$, (P1) reduces to the conventional sum-rate maximization problem in the multi-user MISO systems, which can be tackled by the weighted minimum mean square error (WMMSE) algorithm  \cite{shi2011wmmse}. Specifically, by introducing two auxiliary variables $\boldsymbol{\chi}=\left[\chi_1,\chi_2,\cdots,\chi_K\right]^\top\in\mathbb{C}^{K\times1}$ and $\boldsymbol{\kappa}=\left[\kappa_1,\kappa_2,\cdots,\kappa_K\right]^\top\in\mathbb{C}^{K\times1}$, (P1) can be equivalently transformed into the following optimization problem\vspace{-3pt}
	\begin{subequations}
		\begin{align}
			{\text{(P4-1)}}\quad \underset{\boldsymbol{\chi},\boldsymbol{\kappa},\mathbf{W}}{\max}\quad f_1(\boldsymbol{\chi},\boldsymbol{\kappa},\mathbf{W}) \nonumber \quad \mathrm{s.t.}\quad \eqref{P1a},
		\end{align}
	\end{subequations}
	where $f_1(\boldsymbol{\chi},\boldsymbol{\kappa},\mathbf{W})=\sum_{k=1}^{K}\kappa_ku(\chi_k,\mathbf{W})-\log_2\kappa_k$ and $u(\chi_k,\mathbf{W})=|\chi_k|^2(\sum_{i=1}^{K}\left|\mathbf{h}_k^H\left({\mathbf a}\right)\mathbf{w}_i\right|^2+\sigma^2_n)-\mathrm{Re}\{\chi_k^*\mathbf{h}_k^H\left({\mathbf a}\right)\mathbf{w}_k\}+1$. After the above transformation, the original problem becomes more tractable and can be efficiently solved by updating $\boldsymbol{\chi}$, $\boldsymbol{\kappa}$,  and $\mathbf{W}$ iteratively. Specifically, in the $(l+1)$-th iteration of the WMMSE algorithm, these three variables are calculated based on the following updating rules:\vspace{-6pt}
	\begin{subequations}\label{eqn_WMMSE}
		\begin{align}	&\chi_k^{(l+1)}=\left(\sum_{i=1}^{K}\left|\mathbf{h}_k^H\left({\mathbf a}\right)\mathbf{w}_i^{(l)}\right|^2+\sigma^2_n\right)^{-1}\mathbf{h}_k^H\left({\mathbf a}\right)\mathbf{w}_k^{(l)},\\
	&\kappa_k^{(l+1)}=\left(1-\chi_k^{*(l+1)}\mathbf{h}_k^H\left({\mathbf a}\right)\mathbf{w}_k^{(l)}\right)^{-1},\\
    &\mathbf{w}_k^{(l+1)}=\chi_k^{(l+1)}\kappa_k^{(l+1)}\Big(\mu\boldsymbol{I}_N+\sum_{i=1}^{K}\left|\chi_i^{(l+1)}\right|^2\kappa_i^{(l+1)}\nonumber\\
    &\qquad\quad\qquad\times\mathbf{h}_i\left({\mathbf a}\right)\mathbf{h}_i^H\left({\mathbf a}\right)\Big)^{-1}\mathbf{h}_k^H\left({\mathbf a}\right),
\end{align}
\end{subequations}
where $\mu\ge0$ is the optimal dual variable that ensures $\sum_{k=1}^{K}\left\|\mathbf{w}_k\right\|_2^2\le P$. Note that $\sum_{k=1}^{K}\left\|\mathbf{w}_k\right\|_2^2$ is a monotonically decreasing function w.r.t. $\mu$. Thus, we can find $\mu$ via a bisection search.

\subsubsection{Optimizing $\mathbf{a}$ with Given $\mathbf{W}$}
Next, we optimize the AIV ${\mathbf a}$ for any given transmit precoding matrix $\mathbf{W}$, and (P1) is simplified into the following problem:
\begin{subequations}
	\begin{align}
		{\text{(P4-2)}}\quad &\underset{{\mathbf a}}{\max}\quad R_{\mathrm{sum}}({\mathbf a}),\quad
		\mathrm{s.t.}\quad \eqref{P1b},\,\,\eqref{P1c}.\nonumber
	\end{align}
\end{subequations}
Due to the discrete nature of the AIV ${\mathbf a}$, (P4-2) can be solved by the sequential update algorithms. Specifically, we sequentially update the position of each MA over multiple rounds, each including $N$ iterations. In the $n$-th iteration, we only optimize the position index of the $n$-th MA (i.e., $a_n$), while keeping the positions of the other $(N-1)$ MAs fixed. Let us consider the $n$-th iteration in the $r$-th round and denote by $a_j^{(r)}$ the updated position index of the $j$-th MA in this round, $1\le j \le n-1$. Then, the set of all feasible position indices for optimizing $a_n$ is given by
	\begin{align}
		\mathcal{X}_n^{(r)}&=\left\{a\left|a\in\mathcal{M},\left|a-a_j^{(r)}\right|\ge a_{\min},\forall1\le j \le n-1,\right.\right.\nonumber\\
		&\qquad\,\,\,\,\left.\left|a-a_j^{(r-1)}\right|\ge a_{\min},\forall n+1\le j \le N\right\},\label{eqn_SU_DPVOpt_Set}
	\end{align}
for $1<n<N$. In addition, we set $\mathcal{X}_1^{(r)}=\{a|a\in\mathcal{M},|a-a_j^{(r-1)}|\ge a_{\min},\forall2\le j \le N\}$ and $\mathcal{X}_N^{(r)}=\{a|a\in\mathcal{M},|a-a_j^{(r)}|\ge a_{\min},\forall1\le j \le N-1\}$. Let $\boldsymbol{\hat{a}}=\{a_1^{(r)},\cdots,a_{n-1}^{(r)},a_n,a_{n+1}^{(r-1)},\cdots,a_{N}^{(r-1)}\}$. Then, we can optimize $x_n$ as
\begin{equation}\label{eqn_OptTxPos_n_r}
	a_n^{(r)}=\arg\underset{a_n\in\mathcal{X}_n^{(r)}}{\max}\,\,R_{\mathrm{sum}}(\boldsymbol{\hat{a}}),
	\end{equation}
which can be optimally solved via an enumeration within $\mathcal{X}_n^{(r)}$. Next, we can proceed to update the position of the $(n+1)$-th MA in this round.


Although the AO algorithm can effectively solve (P1), it only ensures local optimality and entails a cubic complexity in the number of MAs ($N$). Moreover, it requires full knowledge of the channel map $\mathcal{H}$. To overcome these limitations, we exploit the spatial correlations of CSI across different positions and develop a more efficient learning-based approach for multi-user cases that avoids explicit channel estimation.\vspace{-6pt}

\subsection{Proposed Optimization Framework with Implicit CSI}
The main idea of our proposed learning-based design is to bypass explicit channel estimation and to solve problem (P1) directly. Specifically, we only measure the channel response vectors from a small set of $S$ fixed sampling points to each of the $K$ users, with $S \ll M$. Similar to the single-user case, we assume a TDD system here to obtain the downlink channel by estimating their uplink counterparts. 
Let the collection of the channel measurements over the $S$ training sampling points to be denoted by
\begin{align}
\mathbf{H}_{\mathrm{probe}}
&=
\begin{bmatrix}
h_{1,1} & h_{1,2} & \cdots & h_{1,S} \\
h_{2,1} & h_{2,2} & \cdots & h_{2,S} \\
\vdots  & \vdots  & \ddots & \vdots  \\
h_{K,1} & h_{K,2} & \cdots & h_{K,S}
\end{bmatrix} \\[0.5em]
&=
\begin{bmatrix}
\mathbf{h}_1, \mathbf{h}_2, \cdots, \mathbf{h}_K
\end{bmatrix}^{\top}
\in \mathbb{C}^{K\times S},\nonumber
\end{align}
\noindent where $\mathbf{h}_k \in \mathbb{C}^S$ denotes the channel vector between the $S$ sampling points and user $k$. It is worth noting that unlike the single-user case, determining the MA positions in the multi-user case requires channel phase information as well due to the presence of inter-user interference. Thus, taking $\mathbf{H}_{\mathrm{probe}}$ as the system input, we aim to learn a mapping function 
\begin{equation}
    g(\cdot): \mathbb{C}^{K \times S} \rightarrow \mathcal{M}^{N},
\end{equation}
which directly predicts the optimal AIV $\mathbf{a}$. Given the input $\mathbf{H}_{\mathrm{probe}}$ and the mapping $g(\cdot)$, the optimization problem for the multi-user scenario can be formulated as  
\begin{align}
\text{(P5)} \quad & \max_{\mathbf{W},\mathbf{a} = g(\mathbf{H}_{\mathrm{probe}})} \quad R_{\mathrm{sum}}(\mathbf{W}, \mathbf{a}) \nonumber\\
\text{s.t.} \quad & \eqref{P1a}, \ \eqref{P1b}, \ \eqref{P1c}.\label{eq:p3_multi} 
\end{align}
Compared with the original problem (P1), solving (P3) eliminates the need for estimation of the full channel maps associated with all users,  and the AIV $\mathbf{a}$ is directly predicted via $\mathbf{H}_{\mathrm{probe}}$. After the positions are predicted by the neural network, the precoding matrix $\mathbf{W}({\mathbf a})$ can be obtained by conducting real-time channel estimation for these predicted positions and following the procedures presented in Section IV-A. However, solving (P3) remains challenging due to the highly complex and implicit nature of $g(\cdot)$. Moreover, the supervised learning adopted in the single-user case becomes ineffective, since only a suboptimal solution to (P1) can be obtained by the AO algorithm presented in Section IV-A.

To address this difficulty, we approximate $g(\cdot)$ using an  attention-based neural network trained to predict optimal MA positions by directly maximizing the system sum-rate in an unsupervised manner. In the offline stage, we generate a large dataset of diverse multi-user channel realizations without the need of labels. The neural network is trained using a composite loss function consisting of a negative sum-rate term and a penalty term associated with the violation of the minimum spacing constraint. In the real-time transmission, the BS measures $\mathbf{H}_{\mathrm{probe}}$, feeds it into the pre-trained DNN, and directly obtains the optimized MA positions.\vspace{-6pt}

\subsection{Model Design}
\begin{figure*}[t]
  \centering
  \includegraphics[width=0.85\linewidth]{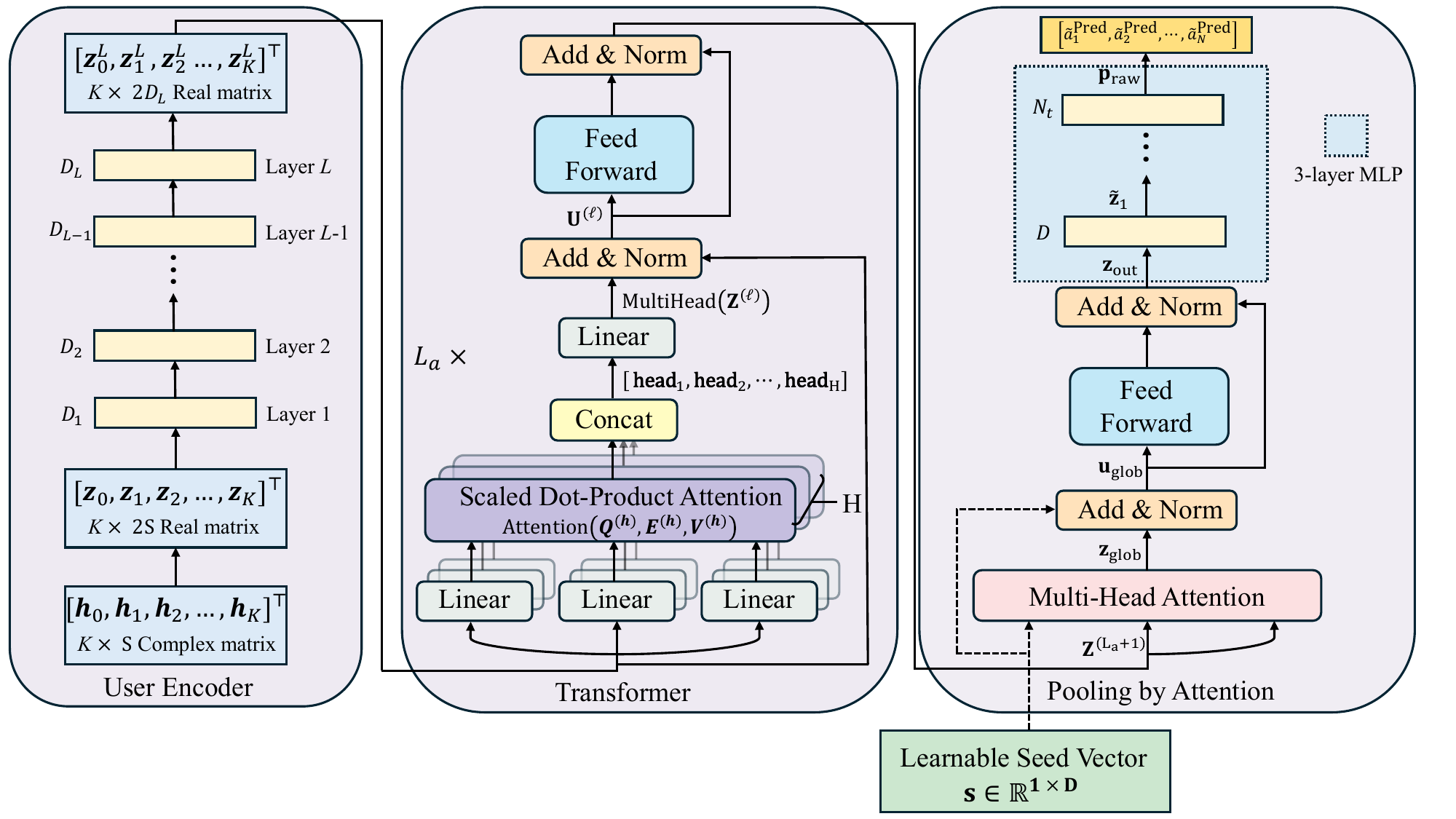}
  \caption{Proposed attention-based architecture in the multi-user case.}
  \label{fig:multiuser model}
  \vspace{-6pt}
  \label{fig:art1}
\end{figure*}
We now present the proposed neural network architecture and the training procedures. The overall architecture is shown in Fig. \ref{fig:art1}.

\subsubsection{User Encoder}
The user encoder takes the channel information $\mathbf{H}_{\mathrm{probe}}$, obtained from the $S$ training sampling points, as input. It then employs an $L$-layer neural network to produce a set of latent features, $\{\mathbf{z}_k^L\}_{k=1}^{K}$. Since standard deep learning frameworks do not support complex-valued operations in general, each row vector $\mathbf{h}^\top_k$ in $\mathbf{H}_{\mathrm{probe}}$ is decomposed into real and imaginary parts and subsequently concatenated as
\begin{equation}
    \mathbf{z}_k^0  
= \big[\operatorname{Re}\{\mathbf{h}^\top_k\},\, \operatorname{Im}\{\mathbf{h}^\top_k\}\big]^\top  
\in \mathbb{R}^{2S \times 1}.
\end{equation}
Given the input feature vector $\mathbf{z}_k^0$, we use $L$ layers of fully connected neural networks, denoted as $\phi_e^l(\cdot)$, $l = 1, 2, \ldots, L$, to produce a high-dimensional feature representation, $\mathbf{z}_k^L$, for user $k$'s channel information. Let $D_l$ denote the dimensionality of  output features at the $l$-th layer. The above process is expressed as
\begin{equation}
    \mathbf{z}_k^{l} = \phi_e^l(\mathbf{z}_k^{l-1}) \in \mathbb{R}^{D_l\times1},\;  k \in {\cal K}.
\end{equation}
At the $l$-th layer, the operation $\phi_e^l(\cdot)$ is carried out as:
\begin{equation}
    \mathbf{z}_k^{l} = \phi_e^l(\mathbf{z}_k^{l-1}) 
    =
    \sigma (\text{LN}(\boldsymbol{W}_{l}\mathbf{z}_k^{l-1} + \mathbf{b}_{l})),
\; 1\leq l \leq L,
\end{equation}
where $\sigma(\cdot)$ denotes the ReLU activation function, and $\boldsymbol{W}_l$ and $\mathbf{b}_l$ are the weight matrix and bias vector for layer $l$, respectively. The layer normalization ($\text{LN}$) is applied to the output of each layer and expressed as
\begin{equation}
    \text{LN}(\mathbf{z}) = \boldsymbol{\gamma} \odot \frac{\mathbf{z} - \boldsymbol{\mu}_\mathcal{L}}{\sqrt{\sigma^2_\mathcal{L} + \varepsilon}} + \boldsymbol{\beta},
\end{equation}
where $\boldsymbol{\mu}_\mathcal{L}\in \mathbb{R}^{D_l \times 1} $ and $\sigma^2_\mathcal{L}\in \mathbb{R}^{D_l \times 1} $ are the mean and variance of the elements in $\mathbf{z}$, respectively; $\boldsymbol{\gamma}\in \mathbb{R}^{D_l \times 1} $ and $\boldsymbol{\beta}\in \mathbb{R}^{D_l \times 1} $ are learnable scaling and shifting parameters, respectively; $\varepsilon$ is a small positive constant introduced for numerical stability (e.g., $1 \times 10^{-5}$ set in our simulation); and $\odot $ denotes the element-wise product. 
After being processed through $L$ layers, the resulting representation vectors, $(\mathbf{z}_1^L, \mathbf{z}_2^L, \ldots, \mathbf{z}_K^L)$, now contain the essential channel features between the users and the BS. These vectors are subsequently fed into an attention module, which is designed to manage the interference among different users. 

\subsubsection{Multi-Head Self-Attention Module}
The output features of the user encoder, i.e., $\left\{\mathbf{z}_k^L\right\}_{k=1}^K$, are then fed into $L_a$ stacked multi-head self-attention blocks that model inter-user interference. We now take one such block as an example to elaborate upon its structure. For clarity, we denote by $\mathbf{Z}^{(\ell)}=[\mathbf{z}^{\ell}_1;\cdots;\mathbf{z}^{\ell}_K]\in\mathbb{R}^{K\times D}$ and $\mathbf{Z}^{(\ell+1)}\in\mathbb{R}^{K\times D}$ as the input and output of the $l$-th block, respectively, $\ell=1,\ldots,L_a$, where $D$ is the hidden dimension of the block .

Multi-head attention allows the model to jointly attend to information from different representation subspaces at different positions, making it well-suited for modeling inter-user couplings. For each multi-head self-attention block, we employ $H$ heads each with a head dimension $d_h$, such that $H \times d_h = D$. Given the input $\mathbf{Z}^{(\ell)} \in \mathbb{R}^{K \times D}$, we first compute the queries ($\mathbf{Q}$), keys ($\mathbf{E}$), and values ($\mathbf{V}$) through the following three linear projections, i.e.,
\begin{align}
\mathbf{Q} &= \mathbf{Z}^{(\ell)} \mathbf{W}^{Q,{(\ell)}} \in \mathbb{R}^{K \times D}, \nonumber\\
\mathbf{E} &= \mathbf{Z}^{(\ell)} \mathbf{W}^{E,{(\ell)}} \in \mathbb{R}^{K \times D}, \label{A}\\
\mathbf{V} &= \mathbf{Z}^{(\ell)} \mathbf{W}^{V,{(\ell)}} \in \mathbb{R}^{K \times D}.\nonumber
\end{align}
where $\mathbf{W}^{Q,{(\ell)}}, \mathbf{W}^{E,{(\ell)}}$ and $\mathbf{W}^{V,{(\ell)}} \in \mathbb{R}^{D \times D}$ are the learnable linear maps for the $l$-th block. 
The resulting matrices in \eqref{A} are then reshaped and split into $H$ parts along the last dimension, denoted as 
\begin{align}
\mathbf{Q} &= [\mathbf{Q}^{(1)}, \mathbf{Q}^{(2)}, \, \cdots , \mathbf{Q}^{(H)}], \nonumber\\
\mathbf{E} &= [\mathbf{E}^{(1)}, \mathbf{E}^{(2)}, \, \cdots, \mathbf{E}^{(H)}], \\
\mathbf{V} &= [\mathbf{V}^{(1)}, \mathbf{V}^{(2)}, \, \cdots , \mathbf{V}^{(H)}].\nonumber
\end{align}
with $\mathbf{Q}^{(h)},\mathbf{E}^{(h)},\mathbf{V}^{(h)} \in \mathbb{R}^{K \times d_h}$. For the $h$-th head, the $\mathbf{Q}^{(h)},\mathbf{E}^{(h)}$ and $\mathbf{V}^{(h)}$ are taken as input, and the output is computed as
\begin{align}
\mathbf{head}_h \;&=\; \mathrm{Attention}(\mathbf{Q}^{(h)}, \mathbf{E}^{(h)}, \mathbf{V}^{(h)}) \notag \\
&=\; \mathrm{softmax}\!\left(\tfrac{\mathbf{Q}^{(h)} (\mathbf{E}^{(h)})^\top}{\sqrt{d_h}}\right) \mathbf{V}^{(h)} 
\;\in\; \mathbb{R}^{K \times d_h}.
\end{align}
where $\mathrm{softmax}(\cdot)$ is applied to each entry of its argument, i.e.,
\begin{equation}
\big(\mathrm{softmax}(\mathbf{S})\big)_{i,j}
= \frac{\exp(S_{i,j})}{\sum_{m=1}^{K}\exp(S_{i,m})},
\end{equation}
where $S_{i,j}$ denotes the $(i,j)$-th entry of $\mathbf{S}$. Then, all the heads are concatenated and followed by an output projection as
\begin{equation}
    \mathrm{MultiHead}(\mathbf{Z}^{(\ell)}) 
= \big[\,\mathbf{head}_1, \mathbf{head}_2, \cdots, \mathbf{head}_H\,\big] \mathbf{W}^{O,(\ell)},
\end{equation}
\noindent with $\mathrm{MultiHead}(\mathbf{Z}^{(\ell)}) \in \mathbb{R}^{K \times D}$ and $ \mathbf{W}^{O,(\ell)} \in \mathbb{R}^{D \times D}$, which serves as an output projection matrix that linearly mixes the concatenated heads.
Then, the blocks adopt the post-norm residual connections, i.e.,
\begin{equation}
    \mathbf{U}^{(\ell)}
=\mathrm{LN}\!\Big(\mathbf{Z}^{(\ell)}+\mathrm{MultiHead}\big(\mathbf{Z}^{(\ell)}\big)\Big)
\in\mathbb{R}^{K\times D},
\end{equation}

\noindent where $\mathrm{LN}(\cdot)$ denotes row-wise layer normalization. Finally, a two-layer position-wise feed-forward network (FFN) is adopted as
\begin{equation}
    \mathrm{FFN}(\mathbf{U^{(\ell)}})=\sigma\!\big(\mathbf{U}^{(\ell)}\mathbf{W}_1^\top+\mathbf{1}\mathbf{b}_1^\top\big)\,\mathbf{W}_2^\top+\mathbf{1}\mathbf{b}_2^\top,
\end{equation}

\noindent where $\mathbf{W}_1\!\in\!\mathbb{R}^{D_{\mathrm{ff}}\times D}$ and $\mathbf{W}_2\!\in\!\mathbb{R}^{D\times D_{\mathrm{ff}}}$ are the weight matrix,  $\mathbf{b}_1\!\in\!\mathbb{R}^{D_{\mathrm{ff}}}$ and $\mathbf{b}_2\!\in\!\mathbb{R}^{D}$are the bias vector, $\mathbf{1}$ is an all-one vector for broadcasting, $\sigma(\cdot)$ is the ReLU activation function, and $D_{\mathrm{ff}}$ denotes the hidden dimension of the FFN. Based on the above, the block output is
\begin{equation}
    \mathbf{Z}^{(\ell+1)}
=\mathrm{LN}\!\Big(\mathbf{U}^{(\ell)}+\mathrm{FFN}\big(\mathbf{U}^{(\ell)}\big)\Big)
\in\mathbb{R}^{K\times D}.
\end{equation}

By stacking the $L_a$ multi-head attention blocks, the multi-head self-attention module outputs $\mathbf{Z}^{(L_a+1)}=[\mathbf{z}^{(L_a+1)}_1,\,\ldots,\,\mathbf{z}^{(L_a+1)}_K]^\top$ for subsequent processing. 

\subsubsection{Global Aggregation Module}
Following the $L_a$ self-attention layers, the user embedding set $\mathbf{Z}^{(L_a+1)} \in \mathbb{R}^{K \times D}$ needs to be aggregated into a fixed-dimensional global representation, which is achieved by adopting pooling-by-multihead-attention (PMA) in this paper. PMA employs a learnable seed vector $\mathbf{s} \in \mathbb{R}^{1 \times D}$ as a single query in a multi-head attention mechanism, where $\mathbf{Z}^{(L_a+1)}$ serves as both key and value matrices.  As such, with the learnable linear maps denoted as $\mathbf{W}^Q_P, \mathbf{W}^K_P$ and $ \mathbf{W}^V_P \in \mathbb{R}^{D \times D}$, we can compute the queries, keys, and values as
\begin{align}
\mathbf{q}_P &= \mathbf{s}\mathbf{W}^Q_P  \in \mathbb{R}^{1 \times D}, \nonumber\\
\mathbf{E}_P &= \mathbf{Z}^{(L_a+1)} \mathbf{W}^E_P \in \mathbb{R}^{K \times D}, \\
\mathbf{V}_P &= \mathbf{Z}^{(L_a+1)} \mathbf{W}^V_P \in \mathbb{R}^{K \times D}.\nonumber
\end{align}
The matrices are then reshaped and split into $H$ parts, denoted as $\mathbf{q}_P^{(h)}
\in \mathbb{R}^{1 \times d_h},\mathbf{E}_P^{(h)} \in
\mathbb{R}^{K \times d_h}$ and $\mathbf{V}_P^{(h)}
 \in \mathbb{R}^{K \times d_h}$ with $H \times d_h = D$. The aggregated representation $\mathbf{z}_{\mathrm{glob}} \in \mathbb{R}^{1 \times D}$ is obtained by concatenating the outputs of $H$ attention heads and following a linear projection $\mathbf{W}_P^O \in \mathbb{R}^{D \times D}$ as
\begin{equation}
\mathbf{z}_{\mathrm{glob}} = [\widetilde{\bf{head}}_1, \dots, \widetilde{\bf{head}}_H] \mathbf{W}_P^O,
\end{equation}
where the output of the $h$-th head is
\begin{equation}
\widetilde{\bf{head}}_h = \mathrm{softmax}\left(\frac{\mathbf{q}_P^{(h)}(\mathbf{E}_P^{(h)})^\top}{\sqrt{d_h}}\right)\mathbf{V}_P^{(h)} \;\in\; \mathbb{R}^{1 \times d_h} .
\end{equation}
Similar to the self-attention blocks, the PMA adopts residual connections and FFN to yield the final global embedding $\mathbf{z}_{\mathrm{out}}$,
\begin{align}
\mathbf{u}_{\mathrm{glob}} &= \mathrm{LN}(\mathbf{s} + \mathbf{z}_{\mathrm{glob}}), \\
\mathbf{z}_{\mathrm{out}} &= \mathrm{LN}(\mathbf{u}_{\mathrm{glob}} + \mathrm{FFN}(\mathbf{u}_{\mathrm{glob}})).
\end{align}
with $\mathbf{u}_{\mathrm{glob}}\in \mathbb{R}^{1 \times D}$ and $\mathbf{z}_{\mathrm{out}} \in \mathbb{R}^{1 \times D}$. The global representation $\mathbf{z}_{\mathrm{out}}$, 
which encodes information from all active users, is projected into a physical antenna coordinate through a regression head $f_{\mathrm{reg}}(\cdot)$ implemented as a three-layer MLP. Specifically, the transformation proceeds as
\begin{align}
&\mathbf{\tilde{z}}_1 = \sigma\big(\mathbf{z}_{\mathrm{out}} \mathbf{W}_1 + \mathbf{b}_1\big),
\; \mathbf{W}_1 \in \mathbb{R}^{D \times D}, \ \mathbf{b}_1 \in \mathbb{R}^{1 \times D}, \\
&\mathbf{\tilde{z}}_2 = \sigma\big(\mathbf{\tilde{z}}_1 \mathbf{W}_2 + \mathbf{b}_2\big),
\; \mathbf{W}_2 \in \mathbb{R}^{D \times D_h}, \ \mathbf{b}_2 \in \mathbb{R}^{1 \times D_h}, \\
&\mathbf{p}_{\mathrm{raw}} = \mathbf{\tilde{z}}_2 \mathbf{W}_3 + \mathbf{b}_3,
\; \mathbf{W}_3 \in \mathbb{R}^{D_h \times N}, \ \mathbf{b}_{3} \in \mathbb{R}^{1 \times N},
\end{align}
where $\mathbf{W}_i$ and $\mathbf{b}_i$ denote the weight matrix and the bias in the $i$-th layer of the MLP, $\sigma(\cdot)$ is the ReLU activation function. Next, the raw predictions $\mathbf{p}_{\mathrm{raw}}$ are further processed to satisfy physical constraints on the antenna positions. For example, the predicted coordinates are restricted to lie within the array size $[0, A]$, for which a sigmoid function is applied as
\begin{equation}
\mathbf{p}_0 = \text{sigmoid}(\mathbf{p}_{\mathrm{raw}}) \cdot A,
\end{equation}
The resulting vector $\mathbf{p}_0 \in \mathbb{R}^{1 \times N}$ represents the estimated positions of the $N$ antenna elements along the array. It is worth noting that the position vector $\mathbf{p}$ remains in a continuous space, which can be simply quantized as the nearest discrete sampling point in the antenna array. \vspace{-6pt}

\subsection{Neural Network Training}
It is worth pointing out that, although the proposed framework eliminates the need for explicit channel estimation during practical deployment, the offline training phase of the neural network still requires full CSI for evaluating the objective function. Moreover, while the WMMSE algorithm is adopted in real-time implementation, its slow convergence speed and difficulties in gradient backpropagation make it less suitable for use in training. To address this issue, this paper employs the regularized zero-forcing (RZF) precoding during the training process, which admits a closed-form expression and thus facilitates efficient gradient-based optimization.  
Specifically, we consider that $T$ independent channel realizations are conducted. In the $t$-th realization, let $\tilde{a}_{t,n}$ denote the predicted index of the sampling point for the $n$-th MA in the $t$-th channel realization by the proposed framework. Given $\{\tilde{a}_{t,n}\}_{n=1}^{N}$, the downlink channel matrix from the BS to the $K$ users is
\begin{equation}\label{channelmap2}
\renewcommand{\arraystretch}{1.4}
\mathbf{H}_{t,p} =
\begin{bmatrix}
    h_{p,1}^{\tilde{a}_{t,1}} & h_{p,1}^{\tilde{a}_{t,2}} & \cdots & h_{p,1}^{\tilde{a}_{t,N}} \\
    h_{p,2}^{\tilde{a}_{t,1}} & h_{p,2}^{\tilde{a}_{t,2}} & \cdots & h_{p,2}^{\tilde{a}_{t,N}} \\
    \vdots & \vdots & \ddots & \vdots \\
    h_{p,K}^{\tilde{a}_{t,1}} & h_{p,K}^{\tilde{a}_{t,2}} & \cdots & h_{p,K}^{\tilde{a}_{t,N}}
\end{bmatrix}
\in \mathbb{C}^{K \times N},
\end{equation}
where $h_{p,k}^{\tilde{a}_{t,n}}$ denotes the complex baseband channel from the predicted antenna position $\tilde{a}_{t,n}$ to user $k$ in the $t$-th realization. For notational brevity, the realization index $t$ is omitted in the sequel without ambiguity. The RZF precoding matrix is given by 
\begin{align}
{\mathbf{W}}_{\mathrm{RZF}}
&= [{\mathbf{w}}_{1,\mathrm{RZF}},\cdots,{\mathbf{w}}_{K,\mathrm{RZF}}] \nonumber\\
&=\; \beta\mathbf{H}_p^H
\big(\mathbf{H}_p \mathbf{H}_p^H + \alpha \mathbf{I}_{K}\big)^{-1},\label{eq:rzf_beamforming}
\end{align}
where  $\alpha = \tfrac{K \sigma_n^2}{P_t}$ is the regularization factor with $\sigma_n^2$ representing the noise power, $\beta$ is a scalar to ensure the maximum power constraint at the BS. Based on this precoding scheme, the sum-rate is given by
\begin{equation} \label{eq:sumrate_final}
    R^{\text{sum}} = \sum_{k=1}^{K} \log_2 \!\left( 1 + \mathrm{SINR}_{k,\mathrm{RZF}} \right),
\end{equation}
where
\begin{equation} \label{eq:sinr_compact}
    \mathrm{SINR}_{k,\mathrm{RZF}} =
    \frac{|\mathbf{h}_{p,k}^H \mathbf{w}_{k,\mathrm{RZF}}|^2}{
    \sum\limits_{j=1,j\neq k}^K{|\mathbf{h}_{p,k}^H \mathbf{w}_{j,\mathrm{RZF}}|^2} + \sigma^2_n}
\end{equation}
denotes the signal-to-interference-plus-noise ratio (SINR) at user $k$ by treating interference as noise.

To ensure that the predicted antenna positions satisfy inter-MA spacing constraints, we introduce a penalty term into the objective function, which penalizes the cases where antenna elements are placed closer than the minimum spacing $d_{\min}$. Mathematically, the penalty term $\mathcal{P}_b$ is defined as
\begin{equation} \label{eq:penalty}
    \mathcal{P}_b = \frac{1}{N}
    \sum_{1\le i < j \le N} \max\big(0, d_{\min} - |\tilde{a}_{i}-\tilde{a}_{j}| \big),
\end{equation}
where $\tilde{a}_i$ is the predicted antenna position for the $i$-th MA.  
By integrating the sum-rate with the penalty function in \eqref{eq:penalty}, the loss function is formulated as
\begin{equation} \label{eq:loss_function_single}
    \mathcal{L} = -R^{\text{sum}} + \eta\, \mathcal{P}_b,
\end{equation}
where $\eta$ is a hyperparameter controlling the trade-off between maximizing the system throughput and satisfying the physical spacing constraint. This joint optimization encourages the model to predict antenna positions that both achieve high system sum-rate and satisfy the minimum spacing requirement. In the exceptional cases where the minimum antenna spacing constraint is still violated, the projection procedures presented in Section III-C-4) can be employed.

\section{Numerical Results}
In this section, we provide numerical results to evaluate the performance of the proposed learning-based solutions. Unless otherwise stated, the simulation settings are as follows. 
The carrier frequency is set to 5 GHz, and thus the wavelength is \(\lambda= 0.06\) meter (m). The number of transmit MAs is $N = 8$, and the length of the linear transmit array is $A = 0.48\,{\text{m}} = 8\lambda$. The minimum distance between any two MAs is set to \(d_{\min}\) = \(\frac{\lambda}{2}\). The distance between any two sampling points is set to $\delta_s=0.1\lambda$; thus, the total number of sampling points is given by $A/\delta_s = M=80$. The BS's transmit power is set to $P_t=35$ dBm. The training sampling points are uniformly distributed within the movement region at the BS. In the offline CSI measurements, the BS receives uplink beacon signals from multiple single-FPA user terminals, whose positions are randomly distributed within a region centered at the BS. Furthermore, we consider the field-response channel model as in \cite{Mei2024}, with the number of receive paths randomly generated within the interval $[3,15]$. Note that in actual downlink communications, these receive paths will become transmit paths. Let $\gamma_i$ denote the channel response coefficient for the $i$-th receive path, which is assumed to follow CSCG distribution with $\gamma_i \sim \mathcal{CN}(0, \beta_0 D^{-\alpha}/L_r)$, where $\alpha= 2.5$ represents the path-loss exponent, $L_r$ denotes the number of receive paths, and $\beta_0 = -52$ dB denotes the path loss at the reference distance of 1 m. The angle of arrival (AoA) for each receive path is assumed to be uniformly distributed within $[0, \pi]$. 

\subsection{Single-User System} 
First, in the single-user case, to obtain the training dataset, we generate $T = 1{,}000{,}000$ random realizations of the user position, the number of receive paths, and the channel response coefficient for each receive path, with the BS-user distance set to 70 m. Both the measurement noise at the BS (in offline training) and the user received noise (in real-time communications) are set to $-$104 dBm, corresponding to a power spectral density of $-164$ dBm/Hz over a bandwidth of $1$ MHz. The transmit power of the user is set to 23 dBm in pilot signal transmission. In each realization, we employ the graph-based algorithm proposed in \cite{Mei2024} to obtain the optimal MA positions as labels for supervised training.

\begin{figure}[t]
    \centering
    \includegraphics[clip, width=0.85\linewidth]{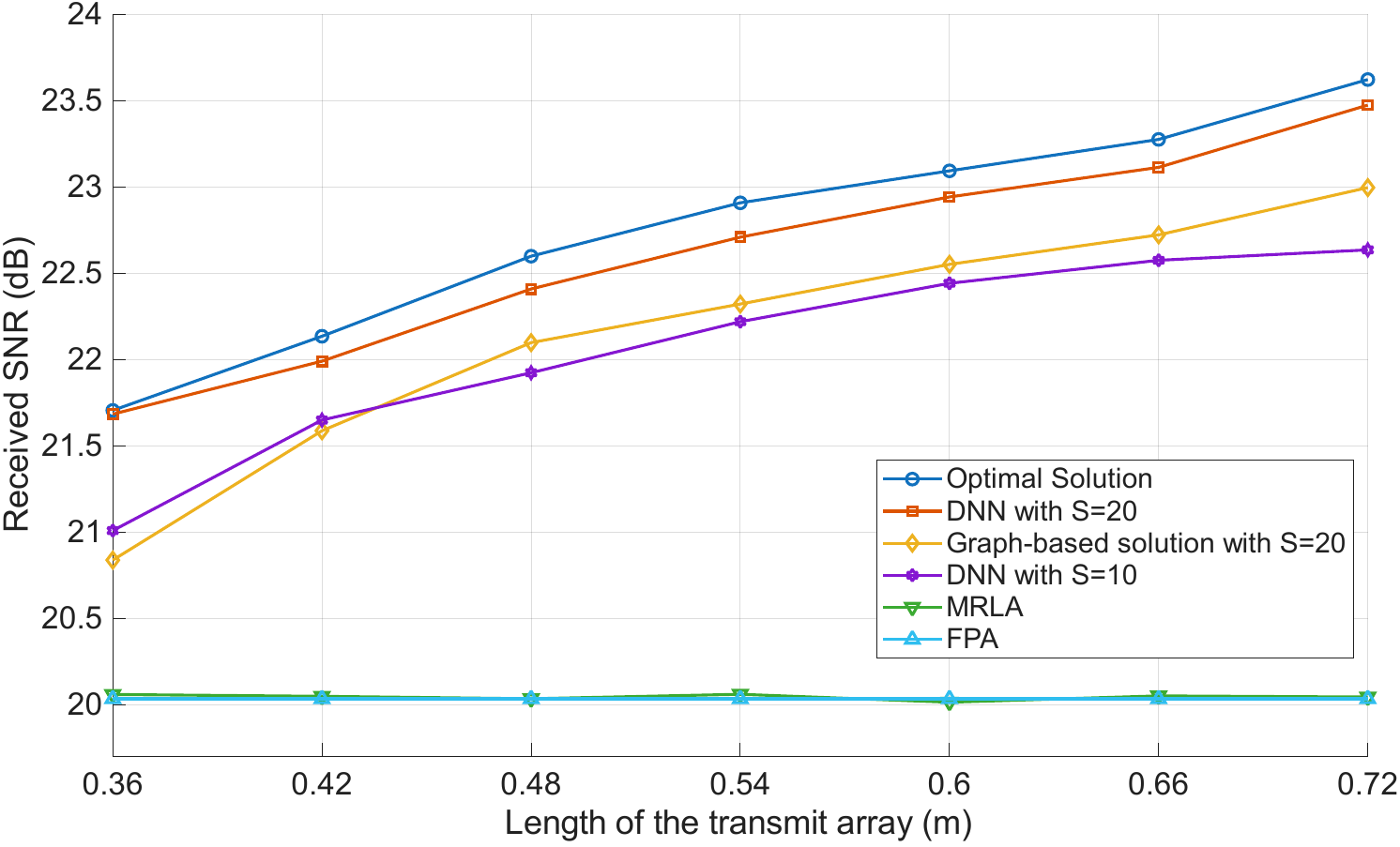}
    \caption{Received SNR versus length of the transmit array.}
    \label{fig:res1}
    \vspace{-9pt}
\end{figure}

Based on the above offline processing, we can obtain the pre-trained DNN. To evaluate its real-time communication performance, we show its average performance over 5000 independent channel realizations. Furthermore, we consider the following benchmark schemes for performance comparison, namely, the optimal graph-based algorithm based on all sampling points \cite{Mei2024}, the conventional half-wavelength-spaced FPAs, and the minimum-redundancy linear arrays (MRLA) \cite{MRLA}. For the MRLA benchmark, the antenna positions are pre-configured offline following a non-uniform linear array geometry to minimize spacing redundancy and maximize spatial resolution \cite{MRLA}. Note that full CSI at all sampling points is required to perform the optimal graph-based algorithm, while the proposed method only requires channel power gains at $S$ training sampling points. Furthermore, we also show the performance of the optimal graph-based algorithm that selects the antenna positions from the same set of $S$ sampling points as the proposed algorithm, which ensures an identical channel estimation overhead in real time.

First, Fig.~\ref{fig:res1} demonstrates the received SNR versus the length of the transmit array, $A$. For our proposed algorithm, we show its performance under $S=$10 and 20, corresponding to 12.5\% and 20\% of the total number of sampling points, respectively. It is observed from Fig.~\ref{fig:res1} that for a small length of the transmit array (e.g., $A=0.36$ m $=$ 6$\lambda$), the proposed DNN-based scheme can achieve a comparable performance to the optimal graph-based solution with $S=20$. Moreover, even with CSI at any $S=10$ sampling points, the performance gap between the proposed scheme and the optimal solution remains within 1 dB. It is also observed that the graph-based algorithm with 20 sampling points yields a worse performance than the proposed scheme with $S=20$. Particularly, as $A=$ 0.42 m, it is even worse than the proposed scheme with $S=10$. This is attributed to the offline training of the proposed scheme that learns the mapping between partial power measurements and optimal antenna positions. The half-wavelength-spaced FPAs and the MRLA benchmark achieve the worst performance among all schemes due to their limited flexibility in antenna repositioning.

Next, we plot the received SNR versus the number of MAs, $N$, in Fig.~\ref{fig:res2}. It is observed that the performance of both the proposed DNN-based algorithm and the benchmark schemes improves with increasing $N$. Particularly, our proposed learning-based approach achieves performance over 95\% of the optimal solution with $S=20$ only, and outperforms the graph-based algorithm with the same set of training sampling points. Furthermore, the performance gap between the DNN algorithm and the optimal solution remains relatively stable as $N$ increases, suggesting that a linear increase in SNR can be achieved by the proposed scheme. Finally, it is observed that the proposed scheme can still outperform the half-wavelength-spaced FPAs and the MRLA benchmark over the whole range of $N$ considered. \vspace{-5pt}

\subsection{Multi-User System}
In the multi-user case, we set the number of fully connected layers $\phi_e^l$ in the user encoder (see Fig.\,\ref{fig:multiuser model}) as $L=3$, while their sizes are set to $2S \times 256$, $256 \times 512$, and $512 \times 512$, respectively. In the multi-head attention blocks, the feature dimension and the number of heads are set to $D= 512$ and $H = 8$, respectively. The hidden feature dimensions in the FFN and aggregation module are set to $D_\mathrm{ff}=2048$ and $D_h=256$, respectively. The number of users is randomly selected in the range from 2 to 6, and their distances from the BS are assumed to be identical as 70 m. Both the measurement noise at the BS (in offline training) and the user received noise (in real-time communications) are set to $-$104 dBm. The transmit power of all users is 23 dBm in pilot signal transmission. The channel with each user is generated following the process mentioned at the beginning of Section V.

\begin{figure}[t]
    \centering
    \includegraphics[clip, width=0.85\linewidth]{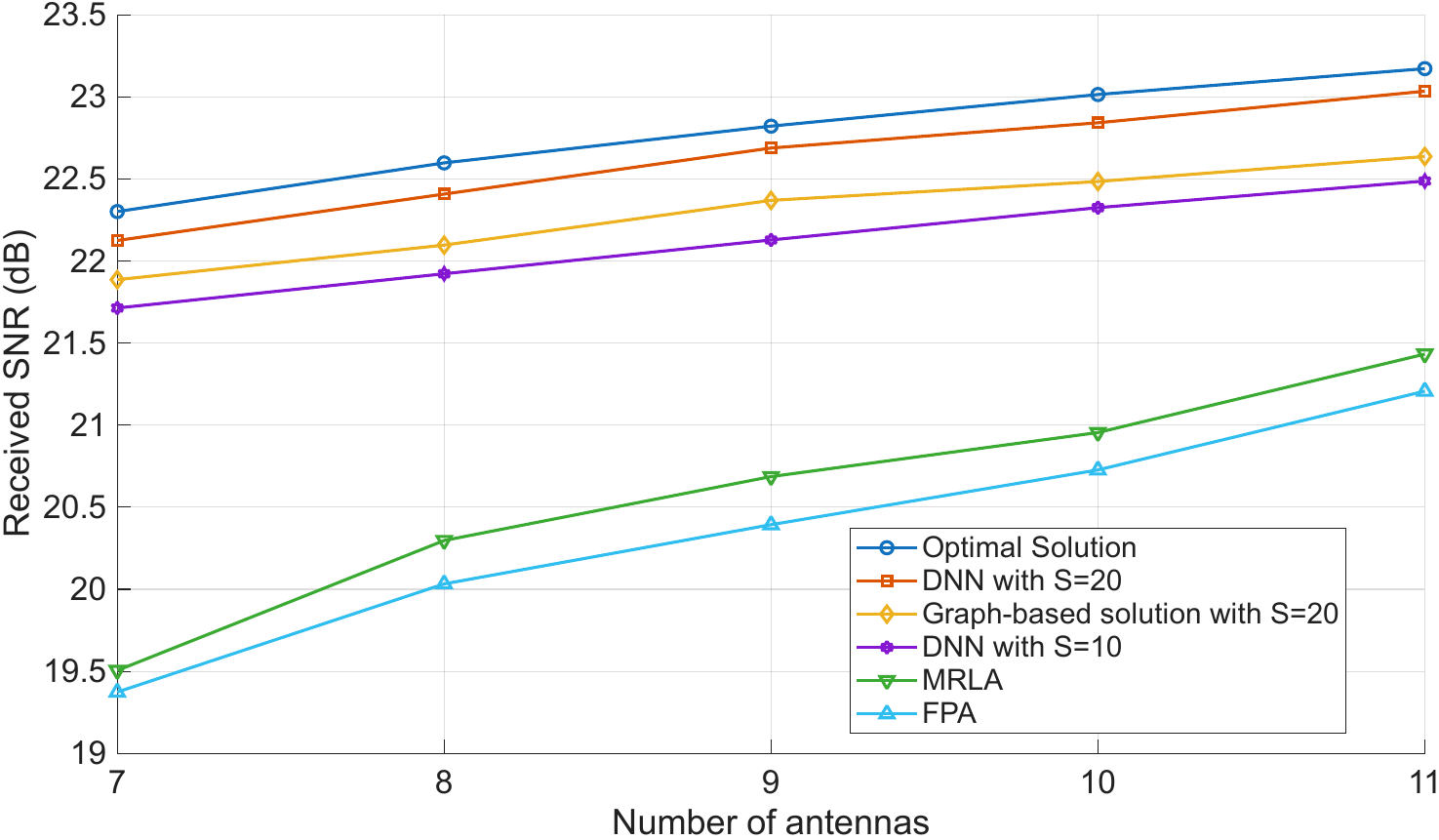}
    \caption{Received SNR versus the number of MAs.}
    \label{fig:res2}
    \vspace{-9pt}
\end{figure}

We implement the proposed network using the deep learning library PyTorch. The neural network is trained using the Adam optimizer with an initial learning rate $10^{-4}$ with a 15-epoch warm-up phase followed by a cosine decay schedule over 100 epochs. At each training epoch, 200,000 training samples are used to compute the gradients in each iteration. We terminate the training process if the loss function does not decrease on the validation data set over 10 consecutive training epochs.

In the actual communications, we show the performance of the proposed method over 1000 independent channel realizations and compare it with the following benchmarks:
\begin{itemize}
    \item \textbf{FPA (Benchmark 1):} The $N$ MAs are deployed symmetrically to (0,0,0) and separated by the minimum distance $d_{min}=\lambda/2$.
    \item \textbf{AO based on full CSI (Benchmark 2):} Given full and perfect CSI, the beamforming vectors and antenna positions are alternately optimized according to the AO method described in Section~\ref{multiUserAO}.  
    \item {\textbf{AO with $S = 20$ (Benchmark 3):}} Based on the CSI at the same sets of training sampling points as the proposed algorithm, the beamforming vectors and antenna positions are alternately optimized following the AO algorithm described in Section~\ref{multiUserAO}.
    \item {\textbf{MRLA (Benchmark 4):}} The positions of the $N$ antennas are pre-configured as an MRLA \cite{MRLA}.
\end{itemize}
\begin{figure}[t]
    \centering
    \includegraphics[clip, width=0.85\linewidth]{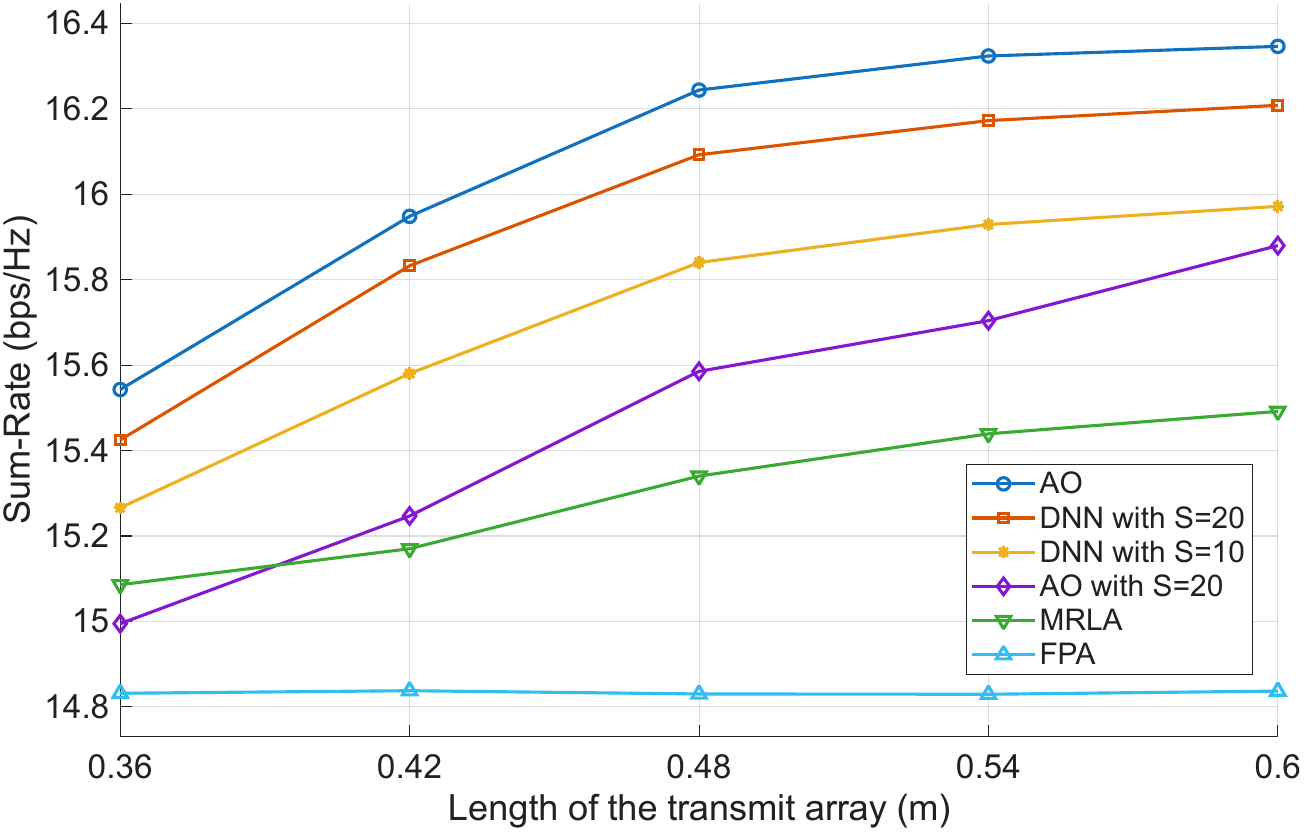}
    \caption{Multi-user sum-rate versus length of the transmit array.}
    \label{fig:res3}
    \vspace{-9pt}
\end{figure}
\begin{figure}[t]
    \centering
    \includegraphics[clip, width=0.85\linewidth]{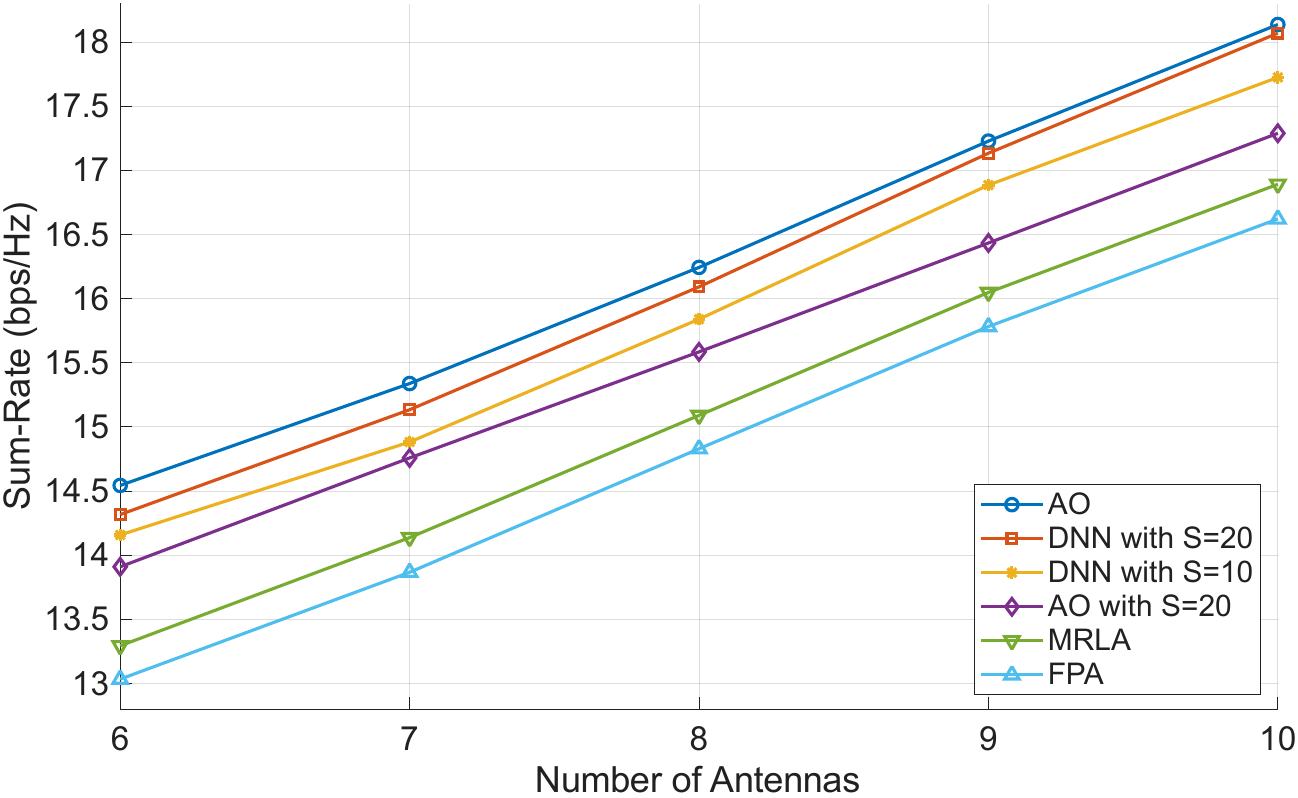}
    \caption{Multi-user sum-rate versus the number of MAs.}
    \label{fig:res4}
    \vspace{-9pt}
\end{figure}

Fig.~\ref{fig:res3} plots the sum-rate versus the transmit array length, $A$, in the multi-user case. As observed from Fig.~\ref{fig:res3}, the proposed attention-based scheme significantly outperforms the half-wavelength-spaced FPAs and the MRLA, and the performance gap increases with $A$ thanks to the enlarged degrees of freedom for antenna position optimization or selection. It is also observed that the performance gap between the proposed scheme (with $S=20$) and the AO algorithm implemented under full CSI is small for all values of $A$ considered. Moreover, if only $S=20$ training sampling points are used, the AO algorithm becomes significantly less competitive and even yields a worse performance than our proposed scheme with $S=10$. Beyond the observations made from Fig.~\ref{fig:res3}, we also find that as $A$ increases, the proposed attention-based method requires significantly less computational time than the AO method. This is because the AO algorithm must explore an increasingly large set of candidate positions as $A$ grows, while the inference/mapping time of our proposed method remains largely unaffected. These results demonstrate the efficiency of the proposed scheme compared to conventional CSI-based approaches in the multi-user setting.
\begin{figure}[t]
    \centering
    \includegraphics[clip, width=0.85\linewidth]{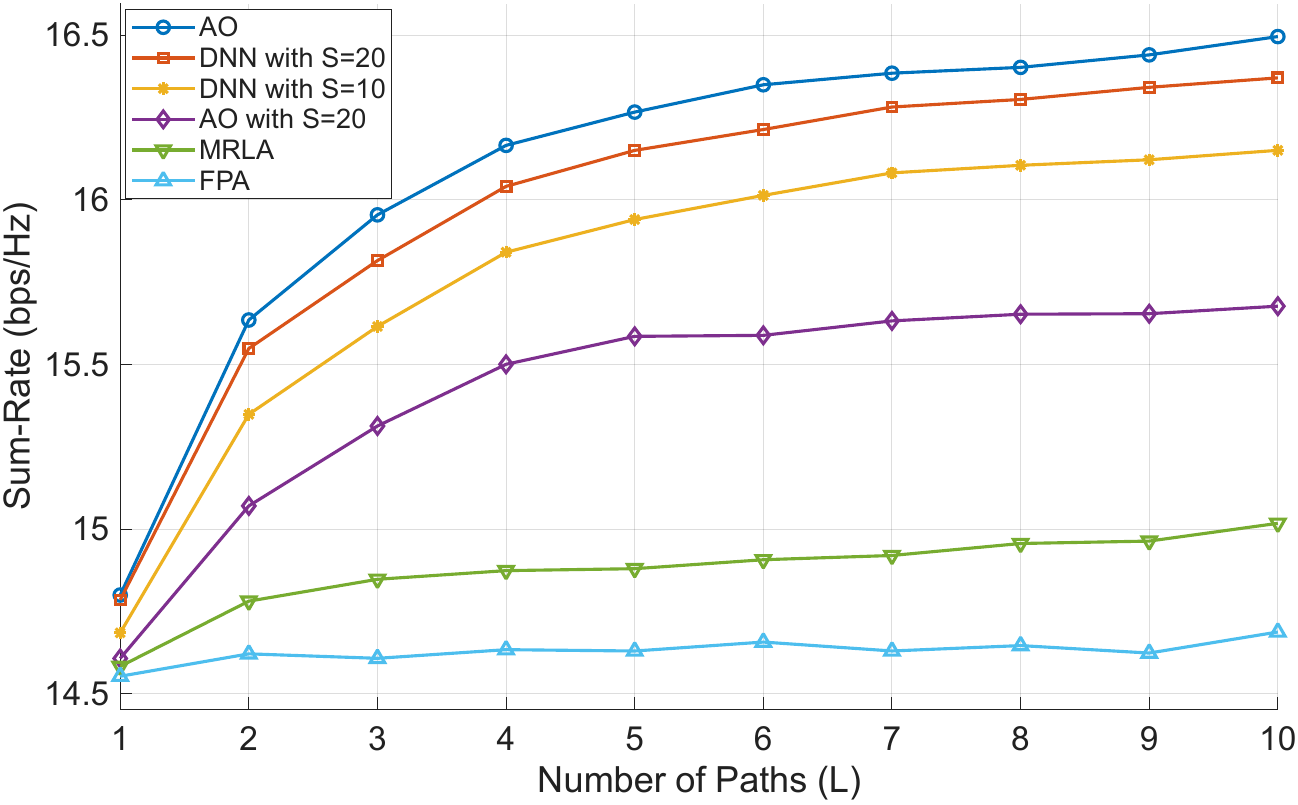}
    \caption{Multi-user sum-rate versus the number of channel paths.}
    \label{fig:res5}
    \vspace{-9pt}
\end{figure}
\begin{figure}[t]
    \centering
    \includegraphics[clip, width=0.85\linewidth]{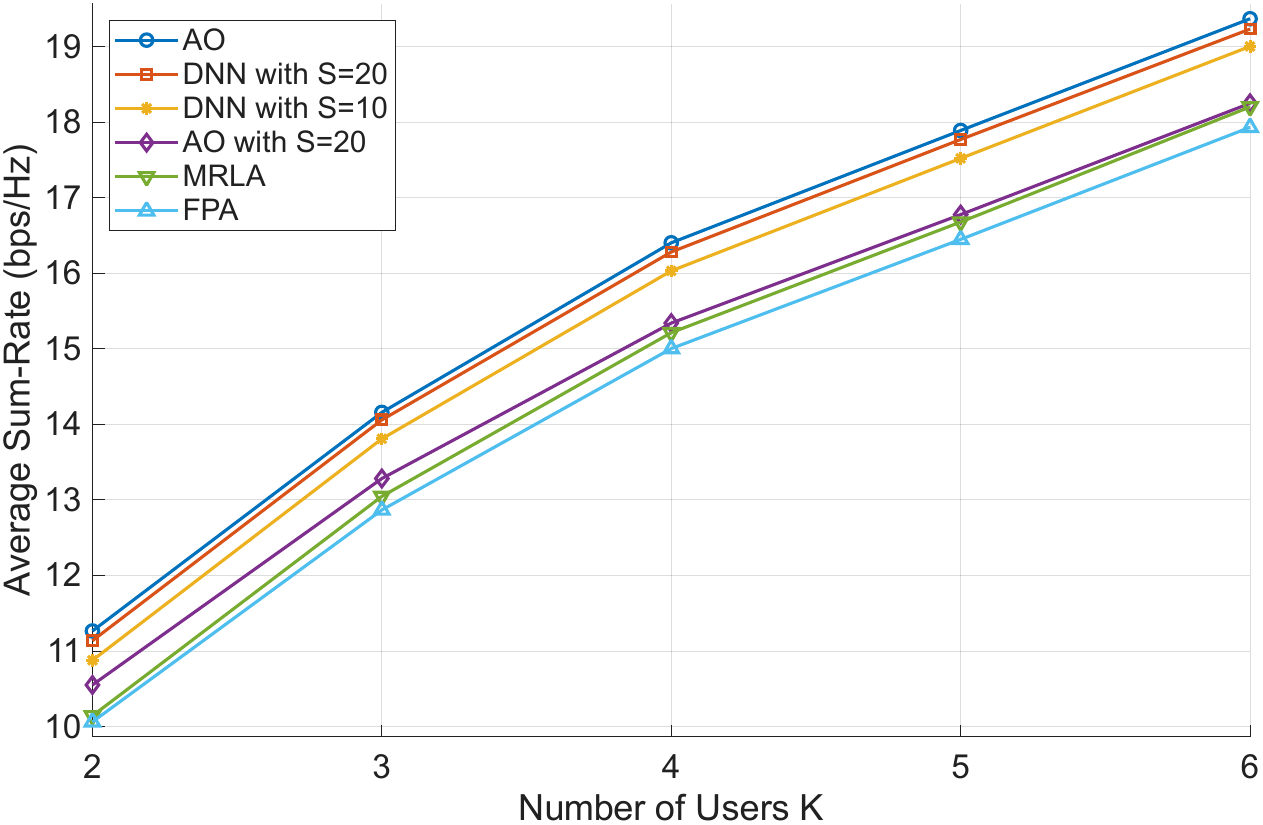}
    \caption{Multi-user sum-rate versus the number of users.}
    \label{fig:res6}
    \vspace{-9pt}
\end{figure}

Fig.~\ref{fig:res4} plots the sum-rate versus the number of MAs, $N$. Similar to the single-user case, the performance of all considered schemes improves as $N$ increases. Notably, the proposed scheme consistently outperforms the AO benchmark under the same CSI input, the MRLA benchmark, and the half-wavelength-spaced FPA benchmark across all antenna configurations. Furthermore, the performance gap between the proposed algorithm and AO based on full CSI narrows as $N$ increases, since the AO algorithm may be more frequently trapped in low-quality local optima as the number of MAs is large. 

Fig. \ref{fig:res5} shows the sum-rate performance of different schemes versus the number of channel paths, where the total channel power gain is uniformly distributed among these paths. It is observed that as the number of paths increases, the performance of all considered schemes (except FPA) improves due to enhanced spatial diversity. The proposed scheme with $S=20$ or $S=10$ consistently outperforms the AO algorithm with $S=20$, the MRLA benchmark, and the FPA benchmark. Particularly, this gap widens as the number of paths increases, indicating that the proposed scheme is especially effective under more complex propagation conditions.

Finally, Fig. \ref{fig:res6} shows the sum-rate performance of the considered schemes versus the number of users, $K$. As expected, it is observed that the rate performance of all considered schemes improves with $K$. Notably, the proposed scheme consistently outperforms both the half-wavelength-spaced FPA and MRLA baselines, as well as the AO algorithm with the same CSI input. Particularly, for a small number of users (e.g., $K = 2,3$), the performance gap between the proposed scheme and the AO algorithm with $S=20$ is small. However, as the number of users increases, this gap becomes more pronounced, since the enlarged user set makes the antenna position optimization more challenging and increases the likelihood that AO becomes trapped in unfavorable local optima.

\section{Conclusion}
This paper addresses the key challenges of channel map estimation and antenna position optimization in MA systems by proposing a learning-based framework that directly infers high-quality MA positions from partial CSI measurements. For the single-user case, we develop an up–down MLP architecture that effectively captures the complex relationship between partial channel gains and optimal antenna positions. We further extend this framework to the more challenging multi-user scenario by designing an attention-based neural network trained in an unsupervised manner. The proposed architecture facilitates efficient inter-user interference mitigation and provides a scalable solution that accommodates varying numbers of users without retraining.
Simulation results show that the proposed method achieves near-optimal performance in the single-user case and outperforms conventional CSI-based AO methods in the multi-user case, while requiring 75\% less CSI. Notably, the performance gains over AO become more pronounced when the antenna movement region is large or the number of users or channel paths increases, since these regimes substantially complicate CSI-driven optimization and make AO more prone to converging to unfavorable local optima.
This work can be extended in several promising directions. For example, it would be interesting to generalize the proposed method to frequency-division duplex (FDD) systems, where uplink CSI may contain latent information about the downlink channel that could be exploited for antenna position optimization. In addition, it is worthwhile to consider more general system settings and antenna architectures, such as wideband MA systems and six-dimensional MA (6DMA) systems \cite{Shao2025,hua20256dma,shao2026tutorial}.

\begingroup
\footnotesize
\printbibliography
\endgroup

\end{document}